\documentclass[final,3p,times]{elsarticle}

\usepackage[colorlinks=true]{hyperref}

\usepackage{amssymb}

\usepackage{lipsum}
\usepackage[labelsep=period,singlelinecheck=off,textformat=period]{caption}
\usepackage{subcaption}
\usepackage{fancyhdr} 
\usepackage{epsfig}
\usepackage{timesmt}
\usepackage{mathtools,nccmath}
\usepackage{mathrsfs}
\usepackage{wrapfig}
\usepackage[noabbrev]{cleveref}
\creflabelformat{equation}{#2\textup{#1}#3}

\usepackage{pdflscape}
\usepackage{afterpage}
\usepackage{float}
\usepackage{url}
\usepackage{enumitem}
\usepackage{accents}
\usepackage{cuted}
\usepackage{dirtytalk}
\usepackage{multirow}
\usepackage{multicol}
\usepackage{booktabs}
\usepackage{array}
\usepackage{makecell}
\usepackage{colortbl}
\usepackage{soul}
\newcolumntype{n}{>{\columncolor{blue!5}}c}

\usepackage{pbox}
\usepackage[draft,backgroundcolor=orange!20]{todonotes}
\usepackage[ruled,commentsnumbered,linesnumbered,boxed]{algorithm2e}
\SetArgSty{textnormal}
\SetKwComment{Comment}{/* }{ */}

\usepackage[flushleft]{threeparttable}
\usepackage{algpseudocode}
\usepackage{setspace}
\usepackage{courier}
\usepackage{psfrag,pstool}

\newcommand{\comment}[1]{}

\Crefname{equation}{Eq.}{Eqs.}
\Crefname{figure}{Fig.}{Figs.}
\Crefname{tabular}{Tab.}{Tabs.}
\crefname{algocf}{alg.}{algs.}
\Crefname{algocf}{Algorithm}{Algorithms}
\Crefname{fct}{Fact}{Facts }

\usepackage{amsmath}
\usepackage{amssymb}
\usepackage{amsthm}
\usepackage{mathrsfs}

\usepackage{bm}
\usepackage{mathtools}
\usepackage{nicefrac}

\theoremstyle{plain}

\theoremstyle{definition}

\theoremstyle{remark}
\newtheorem{rem}{Remark}

\newcommand{\transpose}{^{\mathrm{T}}}
\newcommand{\supertranspose}[2][0.5ex]{\mbox{\rlap{\raisebox{#1}{$\phantom{#2}{}^\mathrm{T}$}}}#2}
\newcommand{\superscript}[1]{\ensuremath{^{\textrm{#1}}}}

\newcommand{\latin}[1]{\textit{#1}}
\renewcommand{\Re}{\mathbb{R}}

\newcommand{\Am}{{\mathbf A}}

\newcommand{\Exp}{\mathbb{E}}

\newcommand{\Fm}{\mathbf{F}}

\newcommand{\g}{G}

\newcommand{\Mm}{{\mathbf M}}

\newcommand{\Ns}{n_\mathrm{s}}

\newcommand{\sm}{\mathbf{s}}

\newcommand{\uu}{{\mathbf u }}

\newcommand{\w}{{\bm{w}}}

\newcommand{\x}{{\mathbf x }}

\newcommand{\y}{{\mathbf y}}

\newcommand{\z}{{\mathbf z}}

\newcommand{\phii}{\boldsymbol{\phi}}
\newcommand{\psii}{\boldsymbol{\psi}}

\newcommand{\thetaa}{\bm{\theta}}

\newcommand{\Ox}{{\Omega_{\mathcal{X}}}}
\newcommand{\Oy}{{\Omega_{\mathcal{Y}}}}
\newcommand{\Oz}{{\Omega_{\mathcal{Z}}}}

\newcommand{\Nx}{{n_{\mathcal{X}}}}
\newcommand{\Ny}{{n_{\mathcal{Y}}}}
\newcommand{\Nz}{{n_{\mathcal{Z}}}}
\newcommand{\prior}{\mathrm{p}_{\mathcal{X}}(\x)}
\newcommand{\priorZ}{\mathrm{p}_{\mathcal{Z}}(\z)}
\newcommand{\posterior}{\mathrm{p}_{\mathcal{X}}(\x \vert \hat{\y})}
\newcommand{\posteriorZ}{\mathrm{p}_{\mathcal{Z}}(\z \vert \hat{\y})}
\newcommand{\likelihood}{\mathrm{p}_{\mathcal{Y}}(\hat{\y} \vert \x )}
\newcommand{\likelihoodd}{\mathrm{p}_{\eta}(\hat{\y} - F(\x))}
\newcommand{\evidence}{\mathrm{p}_{\mathcal{Y}}(\hat{\y})}
\newcommand{\data}{\hat{\y}}
\newcommand{\noise}{\bm{\eta}}

\newcommand{\dm}{D}
\newcommand{\hmm}{H}
\newcommand{\qpdfX}{\mathrm{q}_{\mathcal{X}}}

\newcommand{\ie}{i.e., }

\newcommand{\supth}[1]{\ensuremath{{#1}^{\text{th}}}}

\newcommand{\Tstrut}{\rule{0pt}{2.6ex}}       

\usepackage{pgfplots}
\usepackage{tikz}

\usetikzlibrary{shapes.misc}
\usetikzlibrary{arrows,arrows.meta,calc,backgrounds}

\tikzstyle{block} = [draw,rectangle,thick,minimum height=2em,minimum width=2em]
\tikzstyle{sum} = [draw,circle,inner sep=0mm,minimum size=2mm]
\tikzstyle{connector} = [->,thick]
\tikzstyle{line} = [thick]
\tikzstyle{branch} = [circle,inner sep=0pt,minimum size=1mm,fill=black,draw=black]
\tikzstyle{guide} = []
\tikzset{>=latex}

\DeclareFontFamily{U}{BOONDOX-calo}{\skewchar\font=45 }
\DeclareFontShape{U}{BOONDOX-calo}{m}{n}{
	<-> s*[1.05] BOONDOX-r-calo}{}
\DeclareFontShape{U}{BOONDOX-calo}{b}{n}{
	<-> s*[1.05] BOONDOX-b-calo}{}
\DeclareMathAlphabet{\mathcalboondox}{U}{BOONDOX-calo}{m}{n}
\SetMathAlphabet{\mathcalboondox}{bold}{U}{BOONDOX-calo}{b}{n}
\DeclareMathAlphabet{\mathbcalboondox}{U}{BOONDOX-calo}{b}{n}

\makeatletter
\newcommand{\hathat}[1]{%
	\begingroup%
	\let\macc@kerna\z@%
	\let\macc@kernb\z@%
	\let\macc@nucleus\@empty%
	\hat{\raisebox{.35ex}{\vphantom{\ensuremath{#1}}}\smash{\hat{#1}}}%
	\endgroup%
}
\makeatother

\creflabelformat{equation}{#2(#1)#3}
\crefrangelabelformat{equation}{#3(#1)#4 to #5(#2)#6}

\labelcrefformat{subequation}{#2(#1)#3}
\labelcrefrangeformat{subequation}{#3(#1)#4 to #5(#2)#6}

\renewcommand{\mathbf}[1]{\bm{#1}}

\graphicspath{{./figures/}}

\usepackage{lineno}

\begin{document}
	
	\begin{frontmatter}
		
		
		
		\title{A dimension-reduced variational approach for solving physics-based inverse problems using generative adversarial network priors and normalizing flows}

		
		\author[label1]{Agnimitra Dasgupta}
		\author[label2]{Dhruv V Patel}
		\author[label3,label4]{Deep Ray}
		\author[label5]{Erik A Johnson}
		\author[label1]{Assad A Oberai\corref{cor1}}
		\cortext[cor1]{\ead{aoberai@usc.edu}Corresponding author}

		\affiliation[label1]{organization={{Department of Aerospace \& Mechanical Engineering,
					University of Southern California}},
			city={Los Angeles},
			postcode={90089}, 
			state={California},
			country={USA}}
		
		\affiliation[label2]{organization={{Department of Mechanical Engineering,
					Stanford University}},
			city={Stanford},
			postcode={94305}, 
			state={California},
			country={USA}}
		
		\affiliation[label3]{organization={{Department of Mathematics,
					University of Maryland}},
			city={College Park},
			postcode={20742}, 
			state={Maryland},
			country={USA}}
		
		\affiliation[label4]{organization={{Institute for Physical Science and Technology,
					University of Maryland}},
			city={College Park},
			postcode={20742}, 
			state={Maryland},
			country={USA}}
		
		\affiliation[label5]{organization={{Sonny Astani Department of Civil \& Environmental Engineering,
					University of Southern California}},
			city={Los Angeles},
			postcode={90089}, 
			state={California},
			country={USA}}

		\begin{abstract}
			
			We propose a novel modular inference approach combining two different generative models --- generative adversarial networks (GAN) and normalizing flows --- to approximate the posterior distribution of physics-based Bayesian inverse problems framed in high-dimensional ambient spaces. We dub the proposed framework GAN-Flow. The proposed method leverages the intrinsic dimension reduction and superior sample generation capabilities of GANs to define a low-dimensional data-driven prior distribution. Once a trained GAN-prior is available, the inverse problem is solved entirely in the latent space of the GAN using variational Bayesian inference with normalizing flow-based variational distribution, which approximates low-dimensional posterior distribution by transforming realizations from the low-dimensional latent prior (Gaussian) to corresponding realizations of a low-dimensional variational posterior distribution. The trained GAN generator then maps realizations from this approximate posterior distribution in the latent space back to the high-dimensional ambient space. We also propose a two-stage training strategy for GAN-Flow wherein we train the two generative models sequentially. Thereafter, GAN-Flow can estimate the statistics of posterior-predictive quantities of interest at virtually no additional computational cost. The synergy between the two types of generative models allows us to overcome many challenges associated with the application of Bayesian inference to large-scale inverse problems, chief among which are describing an informative prior and sampling from the high-dimensional posterior.  GAN-Flow does not involve Markov chain Monte Carlo simulation, making it particularly suitable for solving large-scale inverse problems. We demonstrate the efficacy and flexibility of GAN-Flow on various physics-based inverse problems of varying ambient dimensionality and prior knowledge using different types of GANs and normalizing flows. Notably, one of the applications we consider involves a 65,536-dimensional inverse problem of phase retrieval wherein an object is reconstructed from sparse noisy measurements of the magnitude of its Fourier transform.
			
			
		\end{abstract}
		
		
		
		
	\end{frontmatter}
	
	\section{Introduction}\label{sec:introduction}
	
	Inverse problems are useful for determining the causal factors behind an observed phenomenon but remain challenging to solve. Inverse problems are ill-posed and, as such, they may admit multiple or, in the extreme case, no solutions~\cite{calvetti2018inverse}. Moreover, in most practical applications, the forward problem is nonlinear, the inferred quantity is high-dimensional, and measurements are noisy: all of these factors makes it challenging to solve inverse problems. Deterministic approaches to solving inverse problems result in point estimates, i.e., a single solution to the inverse problem at hand, which precludes other possible solutions. In contrast, the Bayesian paradigm treats inverse problems in a stochastic setting, with the \emph{posterior} distribution characterizing all possible solutions to the inverse problem at hand. Using Bayes' rule, the posterior distribution results from updating the \emph{prior} distribution through a \emph{likelihood} function: this process is known as Bayesian inference. The posterior distribution is also useful for quantifying the relative plausibility of different solutions, popularly known as uncertainty quantification. Bayesian inference is attractive because it is philosophically appealing and conceptually simple while giving additional useful information about uncertainty in the solution. However, the application of Bayesian inference poses many practical and computational challenges. On a practical note, selecting a well-informed prior distribution is crucial to the success of Bayesian inference. However, the task of choosing such a distribution capable of accommodating the myriad of variations encountered in practical solution fields is far from straightforward. Even with a carefully designed prior, Bayesian inference remains computationally challenging because only a few cases permit closed-form posterior distributions, \ie those where the prior and likelihood distributions form a conjugate pair. Unfortunately, scenarios where it is possible to leverage Bayesian conjugacy seldom occur, and the posterior distribution must be approximated using appropriate tools.
	
	
	One way of approximating the posterior distribution is through samples drawn from it. Markov chain Monte Carlo (MCMC) methods have been the workhorse of posterior sampling for almost half a century~\cite{robert2011short}. However, the application of MCMC can be challenging on large-scale inverse problems, \ie when the inferred quantity is high-dimensional --- this is popularly known as the `\textit{curse of dimensionality}'. The difficulty manifests as long mixing times and larger auto-correlations between successive samples of Markov chains~\cite{cui2014likelihood,betancourt2017conceptual}. Many notable advancements have been proposed to improve the performance of MCMC methods. Some advanced methods focus on carefully designing the proposal distributions in high dimensions to reduce mixing times~\cite{neal2011mcmc,girolami2011riemann}. Alternatively, some approaches try to reduce the stochastic dimensionality of the inverse problem~\cite{cui2014likelihood,cui2021unified}. Despite these advancements, the application of MCMC to large-scale inverse problems continues to present significant challenges.  
	
	Variational inference is often considered a computationally efficient alternative approach for approximating the posterior distribution~\cite{blei2017variational}. In this approach, a parameterized family of distributions that permit efficient sampling and density evaluations are used to approximate the posterior distribution. The optimal parameters are chosen by minimizing some measure of divergence between the approximate posterior density induced by the adopted distribution family and the true posterior density. Choosing an expressive approximation family for the posterior distribution is critical to the success of variational inference, and doing so is difficult in high dimensions where very little information is available about the shape of the posterior. Recently, transport maps have emerged as a popular choice in this regard~\cite{el2012bayesian}. Instead of approximating the posterior distribution using a parameterized family of distributions, transport maps mold the prior distribution into the posterior distribution. Thus, instead of optimizing the parameters of a family of distributions, the parameters of the transport map must be optimized.  While variational inference has been shown to scale better than MCMC~\cite{el2012bayesian,andrle2021invertible}, the curse of dimensionality is still a challenge. The number of parameters that must be optimized, when approximating posterior distributions, proliferates as the dimensionality of the inverse problem increases. Similarly, the construction of high-dimensional transport maps can be difficult~\cite{brennan2020greedy}. 
	
	More recently, deep learning models capable of carrying out Bayesian inference while meeting or circumventing the challenges posed by Bayesian inference are gaining popularity~\cite{ongie2020deep}. In particular, deep generative models are at the forefront of deep learning-driven Bayesian inference. Also popular are the conditional counterparts of deep generative models trained using supervised data: given realizations from the prior distribution, synthetic measurements are generated using the forward model, and the training data consists of pairs of the prior realizations and corresponding measurements. In such a supervised setting, conditional generative models can be used to obtain realizations from the posterior distribution for any new measurement. Some popular deep generative models are generative adversarial networks (GANs)~\cite{goodfellow2014generative}, normalizing flows~\cite{rezende2015variational,papamakarios2021normalizing,kobyzev2020normalizing}, and variational auto-encoders~\cite{kingma2013auto}. Among them, GANs possess superior sample generation qualities and intrinsic dimension reduction capabilities~\cite{bond2021deep}. As a result, GANs have been used as a data informative priors~\cite{patel2021solution,bohra2022bayesian}. Conditional GANs have also been used to approximate the posterior distribution~\cite{adler2018deep,ray2022efficacy,ray2023solution}. However, GANs remain notoriously difficult to train and susceptible to mode collapse. Moreover, GANs are implicit generative models \ie it is not possible to evaluate point-wise the probability density induced by a GAN. In contrast, variational auto-encoders allow for the computation of a lower bound on point-wise density values. Variational auto-encoders have also been used to perform Bayesian inference~\cite{sohn2015learning,goh2019solving,sahlstrom2023utilizing}, but tend to produce blurry outputs compared to those from GANs. In contrast to GANs and variational auto-encoders, normalizing flows are explicit generative models and allow for point-wise evaluation of the probability density they induce~\cite{dasgupta2023rein}. Normalizing flows utilize invertible neural networks to construct a bijective transformation~\cite{papamakarios2021normalizing,kobyzev2020normalizing}. Therefore, normalizing flows are natural candidates for transport maps and have been used for variational inference~\cite{rezende2015variational,sun2021deep,dasgupta2021uncertainty}.  Conditional normalizing flows have also been used to approximate the posterior distribution in inverse problems~\cite{padmanabha2021solving,ray2022efficacy,ray2023solution}. However, the application of normalizing flows to large-scale inverse problems is challenging: when the inverse problem at hand is high-dimensional, the memory footprint of high-dimensional normalizing flows is so large that training requires access to extraordinary computational resources~\cite{orozco2023adjoint} (for instance, see~\cite{kingma2018glow} where a GLOW normalizing flow model is trained with a mini-batch size of 1 per processing unit which approximately amounts to 40 GPU weeks). Aside from deep generative models, Bayesian neural networks have also been used for Bayesian inference~\cite{lampinen2001bayesian}. Bayesian neural networks implicitly learn to invert measurements by introducing stochasticity in the weights of a neural network. However, Bayesian neural networks have limited capacity due to the approximations made to make the training tractable~\cite{tran2022all}. We note that the supervised datasets necessary to train many of the aforementioned models may not be readily available and computationally expensive to acquire. 
	
	In this work, we propose a modular unsupervised inference framework --- GAN-Flow --- that couples together GANs and normalizing flows to solve large-scale physics-based inverse problems when the only prior information available is a sample from the true but inaccessible prior distribution. GAN-Flow aims to circumvent the challenges faced by generative models when they are used to perform inference in high-dimensional settings by exploiting the respective strengths of the two types of generative models it employs: the dimension reduction capability of GANs, and the efficient variational inference capability of normalizing flows. More specifically, GAN-Flow employs a Wasserstein GAN (WGAN) to learn a data-driven prior distribution which will be useful for Bayesian inference. Further, the WGAN helps reduce the dimensionality of the inverse problem as the generator component of the GAN serves as an injective map from the low-dimensional latent space to the high-dimensional ambient space where the inverse problem is framed.  Thus, GAN-Flow also leverages the dimension reduction offered by GANs. Recent findings suggest that framing inverse problems in lower-dimensional latent spaces may be advantageous~\cite{patel2021gan}. GAN-Flow also utilizes normalizing flows to approximate the posterior distribution in the latent space. Again, the dimension reduction capability of GANs facilitate the construction of simpler normalizing flow models, which now only need to perform variational inference in the lower-dimensional latent space, thereby reducing the memory footprint of normalizing flow models. As a result, GAN-Flow can be used to tackle large-scale inverse problems. 
	The use of a normalizing flow, which serves as a map between the latent prior and the latent posterior, offers one significant advantage: once the normalizing flow is trained, new samples from the latent posterior can be efficiently generated without taking recourse to the computationally expensive forward model. A new sample from the high-dimensional posterior is generated by successive transformations of a sample from the latent space of the GAN using the trained normalizing flow model followed by the trained generator. In this work, we demonstrate the wide applicability and flexibility of the proposed framework on different large-scale linear and nonlinear inverse problems, different synthetic (simple geometrical features and Shepp-Logan phantoms~\cite{toft1996radon}) and real-world (MRI scans of human knees~\cite{knoll2020fastmri,zbontar2018fastmri}) prior distributions, different GAN architectures (self-attention GANs~\cite{zhang2019self} and GANs that progressively grow~\cite{karras2018progressive}), and normalizing flows with different invertible neural network architectures (planar~\cite{rezende2015variational} and affine-coupling flows~\cite{dinh2014nice,dinh2016density}). 
	
	We must mention a growing body of work that has developed deep learning-based Bayesian inference frameworks with some dimension reduction component. For instance, this work was directly inspired by \cite{patel2021gan,patel2021solution}, where GANs are used to learn prior distributions and MCMC methods are used to sample from its posterior. \citet{patel2021gan} used the Hamiltonian MCMC, whereas \citet{bohra2022bayesian} used the Metropolis-adjusted Langevin MCMC. However, posterior sampling using an MCMC-based method will fail when the latent space dimensionality continues to be high (for example see \Cref{subsec:phase}). Additionally, it is non-trivial to ascertain the convergence of MCMC chains. In contrast, one can easily gauge the convergence of the latent posterior induced by the normalizing flow model by tracking the loss function used to train the normalizing flow model. Moreover, for a fixed compute budget (as defined by the number of forward problem solves), training the normalizing flow model requires less compute wall times since it is possible to train it using mini-batches. Similarly, sampling is also embarrassingly parallelizable. Whereas MCMC is inherently sequential, and samples are obtained iteratively. 
	Bayesian inference approaches that consist of a generative prior coupled with a way of sampling from its posterior are widely known as \emph{modular} Bayesian approaches. GAN-Flow is also a modular approach in that sense, but distinct because it uses a GAN-based prior and a variational posterior induced by a normalizing flow. There are also several works where normalizing flows are constructed in lower-dimensional spaces and subsequently used to solve inverse problems. Some tools that have been used to derive or learn the injective map include principal component analysis \cite{cramer2022principal}, isometric auto-encoders \cite{cramer2022nonlinear} and injective neural networks \cite{kothari2021trumpets}. \citet{brehmer2020flows} also explore padding the low dimensional latent variable with zeroes to increase dimensionality. Other approaches simultaneously learn the dimension reduction map and normalizing flow~\cite{brehmer2020flows,cramer2022principal,cramer2022nonlinear,kothari2021trumpets}.  GAN-Flow is different since it uses a GAN to approximate the prior density, and the generator of the  WGAN serves as the injective map. 
	
	\subsection{Summary of contributions}
	In summary, the novel contributions of this work are as follows:
	\begin{enumerate}
		\item We introduce GAN-Flow, a novel unsupervised modular Bayesian inference framework, that combines two types of generative models --- GANs and normalizing flows --- and exploits their respective strengths. 
		\item We develop a two-stage strategy to train each sub-component of the GAN-Flow. First, the GAN is trained using \latin{a priori} available samples from the prior distribution. Then, the normalizing flow model is used to perform variational Bayesian inference in the low-dimensional latent space of the GAN for efficient posterior approximation. 
		\item We demonstrate the efficacy of GAN-Flow on three large-scale physics-based inverse problems involving both synthetic and real-world data. We consider three inverse problems --- inferring initial conditions in a heat conduction problem, an inverse Radon transform problem wherein an object is recovered from its sinogram, and a phase retrieval problem wherein an object is recovered form the magnitude of its Fourier transform. Where possible we compare GAN-Flow with Monte Carlo simulation and an inference approach previously proposed by \citet{patel2021solution} that also utilizes a WGAN-GP prior.     
		\item We also show that GAN-Flow is a flexible framework that can utilize various types of GAN and normalizing flow models. For the various problems we consider, we use different GAN models which include generators with self-attention units and GANs that are progressively grown.  We also show that GAN-Flow can accommodate different types of normalizing flows such as planar flows and affine-coupling flows.  
	\end{enumerate}
	The remainder of this paper is organized as follows. \Cref{sec:background} sets up the problem of interest, and provides a brief background on variational Bayesian inference, GANs and normalizing flows. We introduce GAN-Flow in \Cref{sec:proposedapproach} and discuss the training of its two sub-components. In \Cref{sec:results}, we apply GAN-Flow to solve three large-scale inverse problems. Finally, we draw conclusions in \Cref{sec:conclusion}.

	\section{Background} \label{sec:background}
	\subsection{Problem setup and Bayesian inference}
	Consider the random vectors $\x \in \Ox \subseteq \Re^{\Nx}$ and $\y \in \Oy \subseteq \Re^{\Ny}$ related by the forward model $F : \Ox \to \Oy$ such that $\y = F(\x)$. Herein, $\x$, $\Ox$ and $\Nx$ are called the \emph{ambient} variable, space and dimension, respectively. The inference of $\x$ from a noisy measurement vector $\data$ (a noisy realization of $\y$) constitutes an inverse problem. Given a likelihood function $\likelihood$, Bayes' rule is used to update prior belief about $\x$, characterized through the prior probability density function $\prior$, as follows:
	\begin{equation}\label{eq:Bayes}
		\posterior = \frac{\likelihood \; \prior}{\evidence},
	\end{equation}
	where $\posterior$ is the posterior distribution and $\evidence$ is the evidence or marginal likelihood. When measurements are corrupted by an additive noise $\noise$, distributed according to $\mathrm{p}_{\eta}$, the measurement model $\data = \y + \noise$ leads to the likelihood function $\likelihood = \likelihoodd$ in \Cref{eq:Bayes}. The posterior distribution is useful for computing posterior-predictive statistics of any desired quantity of interest, herein denoted as $\ell(\x)$. For instance, the posterior mean of $\ell(\x)$ can be computed as follows:
	\begin{equation}\label{eq:poststat}
		\Exp_{\x \sim \posterior} \big[ \ell(\x) \big] = \int_{\Ox} \ell(\x) \posterior \;\mathrm{d}\x 
	\end{equation}
	Typically, the integral in \Cref{eq:poststat} is high-dimensional and intractable for practically interesting problems, and must be approximated using Monte Carlo methods, which requires samples from the posterior distribution. Given a sample of size $\Ns$, the Monte Carlo approximation to \Cref{eq:poststat} is given as:
	\begin{equation}
		\Exp_{\x \sim \posterior} \big[ \ell(\x) \big] \approx \frac{1}{\Ns} \sum_{i=1}^{\Ns} \ell(\x^{(i)}),
	\end{equation}
	where $\x^{(i)}$ is the \supth{i} realization of $\x$ drawn from the posterior distribution. In this work, we propose the novel inference framework GAN-Flow, which is efficient at sampling the posterior distribution and, ultimately, estimating the statistics of posterior-predictive quantities. GAN-Flow is a hybrid of two types of generative models: generative adversarial networks (GANs) and normalizing flows. 
	
	\subsection{Generative Adversarial Networks} \label{subsec:GANs}
	Generative adversarial networks \cite{goodfellow2014generative} are generative models consisting of two sub-networks: a generator and a discriminator (also known as the critic). GANs are trained \emph{adversarially}: the generator tries to deceive the discriminator while the discriminator tries to distinguish between `\emph{fake}' samples generated from the generator and `\emph{true}' samples available from the target distribution. The generator and critic play an adversarial `\emph{game}' between them with the ultimate goal of generating new realizations from an underlying distribution, the prior distribution $\prior$ in this case.  Let the generator network $G$, parameterized by $\thetaa$, map the \emph{latent} variable $\z \in \Oz \subseteq \Re^{\Nz}$ to the target variable $\x$, \ie $G(\cdot, \thetaa) : \Oz \to \Ox$. Herein, we refer to $\z$, $\Oz$ and $\Nz$ as the \emph{latent} variable, space and dimension, respectively. Typically, $\z$ is sampled from a simple distribution $\priorZ$, like the multivariate standard normal distribution.  Moreover, the latent dimension $\Nz$ is typically chosen to be much smaller than the ambient dimension $\Nx$, \ie $\Nz \ll \Nx$. Thus, GANs are endowed with dimension reduction capabilities and the generator $G$ serves as a map from the low-dimensional latent space to the high-dimensional ambient space. On the other hand, the discriminator $\dm$, parameterized by $\phii$ such that $\dm(\cdot, \phii) : \Ox \to \Re$, tries to differentiate between realizations drawn from $\prior$ and those generated by the generator. 
	
	The parameters $\thetaa$ and $\phii$ of the generator and the discriminator networks, respectively, are obtained through the min-max optimization of an appropriate loss function, say $\mathcal{L}_{\mathrm{GAN}}$, \ie
	\begin{equation}\label{eq:GAN-minmax}
		(\thetaa^\ast, \phii^\ast) = \arg\min_{\thetaa} \Big( \arg\max_{\phii} \mathcal{L}_{\mathrm{GAN}}(\thetaa, \phii) \Big). 
	\end{equation}
	Different types of GANs will use different loss functions $\mathcal{L}_{\mathrm{GAN}}$; interested readers may refer to~\cite{hong2019generative,yi2019generative,jabbar2021survey} for an overview. It is important to note that training a GAN requires realizations of $\prior$, therefore, we assume that $n_{\mathrm{data}}$ independent and identically distributed (iid) realizations of $\x$ from $\prior$ are available, which we herein denote using  $\mathcal{S} = \left\{ \x^{(i)} \right\}_{i=1}^{n_\mathrm{data}}$ and refer to $\mathcal{S}$ as the prior dataset. GAN-Flow uses the provided prior dataset to derive a data-driven informative prior. All that's required is a mechanism to sample from the GAN prior. 
	

	\subsection{Variational Bayesian inference}\label{subsec:VBI}
	
	MCMC methods approximate the posterior distribution using  correlated realizations of $\x$ that are sampled from an ergodic Markov chain with the stationary distribution $\posterior$. In contrast, variational Bayesian inference methods attempt to approximate the posterior probability distribution~\cite{blei2017variational}.  Variational Bayesian inference starts with a family of distributions $\qpdfX(\x; \psii)$ parameterized by $\psii$. The optimal parameter vector $\psii^\ast$ is determined by minimizing some divergence measure $d$ between $\qpdfX(\x; \psii)$ and $\posterior$:
	\begin{equation}\label{eq:VBI}
		\psii^\ast = \arg \min_{\psii} d \big( \qpdfX(\x; \psii) \Vert \posterior \big).
	\end{equation}
	The reverse Kullback-Leibler (KL) divergence is a popular choice for $d$ but other divergence measures have also been used~\cite{dhaka2021challenges}. Thus, variational Bayesian inference converts the problem of posterior sampling into an equivalent optimization problem. Once $\psii^\ast$ has been determined, $\qpdfX(\x; \psii^\ast)$ serves as an approximation to $\posterior$ and can be repeatedly sampled without additional likelihood evaluations to obtain as many posterior samples as required --- unlike MCMC-based methods. As a result, variational Bayesian inference offers a computationally efficient alternative to MCMC sampling in many cases. The performance of variational Bayesian inference relies on the \emph{a priori} chosen parameterized family of distribution $\qpdfX(\cdot; \psii)$ being capable of approximating the posterior distributions, which can have a complex shape. This approximation may be difficult to achieve using standard distribution families like mixture models. Moreover, the computational effort of the optimization problem in \Cref{eq:VBI} increases as the dimension of $\psii$ increases, which is expected to happen as the ambient dimensionality of the inverse problem grows.

	
	An alternative approach to explicitly working with a family of distributions is to define a \emph{pushforward} map that can induce a good approximation to the posterior distribution. Let $\hmm(\cdot; \psii) : \Ox \to \Ox$ denote a bijective and differentiable map (also known as a diffeomorphism) that is parameterized by $\psii$, and let $\hmm_{\#}\mathrm{p}_{\mathcal{X}}(\x; \psii)$ denote the pushforward of the prior distribution $\prior$. Then 
	\begin{equation}\label{eq:change_of_vars1}
		\hmm_{\#}\mathrm{p}_{\mathcal{X}}(\x; \psii)=  \prior \lvert \det \nabla_\x \hmm(\x; \psii) \rvert^{-1},
	\end{equation}
	as a result of change of variables, where $\det \nabla_\x \hmm(\x; \psii)$ is the Jacobian determinant of the pushforward map $\hmm(\cdot; \psii)$.  Therefore, one way of approximating the posterior distribution is to use a flexible diffeomorphism such that the pushforward distribution $\hmm_{\#}\mathrm{p}_{\mathcal{X}}(\x; \psii)$ is close (in some sense) to the posterior distribution. However, the successful application of \Cref{eq:change_of_vars1} requires that the Jacobian determinant be easily computable. Many techniques, such as polynomial approximation and radial basis functions, can be used to construct diffeomorphisms that permit efficient Jacobian determinant computations~\cite{marzouk2016introduction}. More recently, normalizing flows~\cite{kobyzev2020normalizing,papamakarios2021normalizing} have emerged as an efficient tool to construct high-dimensional diffeomorphisms.

	\subsection{Normalizing flows}\label{subsec:NFs}
	Normalizing flows are a class of generative models that uses invertible neural networks to construct diffeomorphisms. Normalizing flows are constructed in a manner that facilitates efficient computation of the Jacobian determinant. In practice, the inference map $\hmm$ is constructed by stacking together multiple, say, $n_\mathrm{f}$ invertible layers, which makes  
	\begin{equation}\label{eq:flow_composition}
		\hmm(\x) = \hmm_{[n_\mathrm{f}]} ( \hmm_{[n_\mathrm{f}-1]} ( \cdots \hmm_{[1]}(\x))).
	\end{equation}
	Individual bijections $\hmm_{[k]}$ are called flows and the composition $\hmm$ is a normalizing flow. $\psii$, which we intentionally suppress in \Cref{eq:flow_composition} and herein, denotes the parameters of all flow layers taken collectively. Note that the composition of bijective functions is also a bijective function, and the Jacobian determinant of which can be computed as:
	\begin{equation}
		\det \nabla_{\x} \hmm(\x) = \prod_{k=1}^{n_\mathrm{f}} 	\det \nabla_{\x_{[k-1]}} \hmm_{[k]}(\x_{[k-1]}),
	\end{equation}
	where $\x_{[k]} =  \hmm_{[k]}(\x_{[k-1]})$ and $\x_{[0]} = \x$. 
	
	Many different types of invertible architectures exist that define bijections for which the Jacobian determinant is easily computable; see \cite{kobyzev2020normalizing} for a recent review. In this work, we use two types of invertible architecture.
	
	\subsubsection{Planar flows}
	\citet{rezende2015variational} proposed an invertible neural network architecture based on planar transformations that apply the following perturbation to the input $\x_{[k-1]}$ in \supth{k} flow layer $\hmm_{[k]}$:
	\begin{equation}
		\hmm_{[k]}(\x_{[k-1]})  = \x_{[k-1]} + \uu_{[k]} \cdot S \left( \w_{[k]}\transpose \x_{[k-1]} + b_{[k]} \right), 
	\end{equation}
	where $\psii_{[k]} = \{ \uu_{[k]} \in \Re^d, \w_{[k]} \in \Re^d, b_{[k]} \in \Re \}$ are the parameters of $\hmm_{[k]}$, and $S: \Re \to \Re$ is a nonlinear activation function with derivative $S^\prime$. The taxonomy `planar' derives itself from the fact the perturbation introduced to $\x_{[k-1]}$ is normal to the hyper-plane $\w_{[k]}\transpose \x_{[k-1]} + b_{[k]} = 0$. The Jacobian determinant of $\hmm_{[k]}$ is:
	\begin{equation}\label{eq:jacdetfk}
		\big|\det \nabla_{\x_{[k-1]}}\hmm_{[k]}(\x_{[k-1]}) \big| = \left| 1 + S^\prime(\w_{[k]}\transpose \x_{[k-1]} + b_{[k]})  \uu_{[k]}\transpose \w_{[k]}  \right|. 
	\end{equation}
	Moreover, $\w_{[k]}\transpose \uu_{[k]} \geq -1$ is a sufficient condition for $\hmm_{[k]}$ to be invertible when $S$ is the hyperbolic tangent function~\cite{rezende2015variational}, which is what we use in this work.  
	
	\subsubsection{Affine-coupling flows}
	\citet{dinh2014nice} introduced coupling flows, of which affine-coupling is a specific type. Let $\x^{a}_{[k-1]}$ and $\x^{b}_{[k-1]}$ be two disjoint partitions of the input vector $\x_{[k-1]}$, formed by randomly sampling components of $\x_{[k-1]}$, then the coupling flow layer $\hmm_{[k]}$ applies the following transformations to its input $\x_{[k-1]}$:
	\begin{equation}\label{eq:nvp_split}
		\x^{b}_{[k]} =  \x^{b}_{[k-1]} \odot \exp\left[S_1(\x^{a}_{[k-1]})\right] + T_1(\x^{a}_{[k-1]}) \text{ and } 
		\x^{a}_{[k]} =  \x^{a}_{[k-1]} \odot \exp\left[S_2(\x^{b}_{[k]})\right] + T_2(\x^{b}_{[k]}),
	\end{equation}
	where $\x_{[k]}\transpose = \left[ \supertranspose{\x^{a}}_{[k]}, \supertranspose{\x^{b}}_{[k]} \right]$, $S_1$ and $S_2$ are known as scale networks, $T_1$ and $T_2$ are known as shift networks, and $\odot$ denotes the Hadamard product. The scale and shift networks are modeled using deep neural networks that preserve the dimensionality of their respective inputs. A coupling layer constrains the Jacobian to be upper triangular~\cite{dinh2014nice,dinh2016density}. The determinant of the Jacobian is~\cite{dinh2016density}:
	\begin{equation}
		\big|\det \nabla_{\x_{[k-1]}}\hmm_{[k]}(\x_{[k-1]}) \big| = \left( \exp\left[\sum_{j} \left\{S_1(\x^{a}_{[k-1]})\right\}_j  \right]\right) \left( \exp \left[ \sum_{j} \left\{S_2(\x^{c}_{[k-1]})\right\}_j \right]\right),
	\end{equation}
	where $\left\{ \cdot \right\}_j$ denotes the \supth{j} component of a vector, and $S_1(\x^{a}_{[k-1]})$ and $S_2(\x^{c}_{[k-1]})$ are outputs from the scale networks $S_1$ and $S_2$, respectively. 

	\section{Bayesian inference using GAN-Flow} \label{sec:proposedapproach}
	
	Bayesian inference is useful for solving statistical inverse problems, but its practical application to large-scale inverse problems is far from straightforward. First, it is important to recognize that the quality of inference depends on the prior~\cite{gelman2013philosophy}, more so when there is paucity of data. Simple parametric priors derived from tractable distributions are not useful for describing complex entities such as brain scans, thermal conductivity fields, and the matrix of a composite material; recent recognition of this fact has fostered efforts to develop physics-informed data-driven priors  \cite{meng2022learning,patel2021solution}.  Second, posterior sampling using MCMC methods is difficult in high-dimensional spaces, \ie when $\Nx$ is large. In high-dimensional spaces, Markov chains tend to take a long time before they can reach a `\textit{steady state}', and assessing the convergence of Markov chains is also difficult. Third, MCMC sampling involves repeated evaluations of the likelihood function, which means that the underlying physics-based forward model must be evaluated during sampling and that the cost of obtaining new samples will scale linearly with the cost of forward model evaluations; this is undesirable. Thus, MCMC sampling from high-dimensional posteriors continues to be a challenging and computationally intensive task,  which has been a major deterrent to the practical application of Bayesian inference to large-scale inverse problems. GAN-Flow attempts to circumvent these issues by coupling together two types of deep generative models --- generative adversarial networks and normalizing flows.

	\subsection{Overview of GAN-Flow}
\begin{figure}
		\centering
		\includegraphics[width=\textwidth]{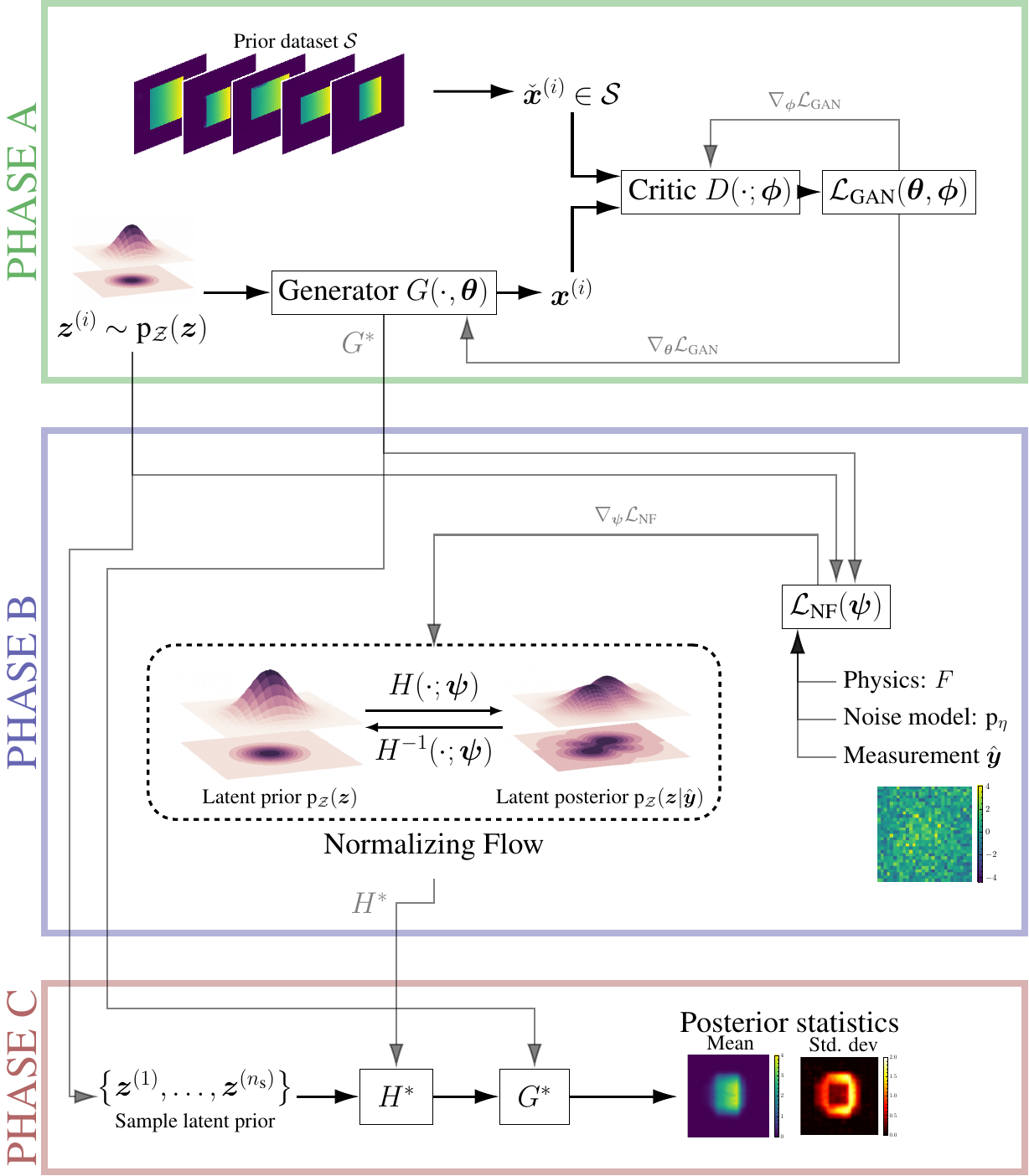}
		\caption{Schematic diagram of the proposed GAN-Flow framework for solving physics-based inverse problems. PHASE A involves the training of a WGAN-GP model with training samples from the prior distribution. In PHASE B, the trained generator $G^\ast$, the physics model $F$, the noise model $\mathrm{p}_\eta$ and the measurements $\data$ are used to train the normalizing flow map $\hmm$. PHASE C corresponds to posterior sampling that is achieved by using the trained normalizing flow map $\hmm^\ast$  to transform realizations from the latent prior into realizations from the latent posterior, which are then passed through the trained generator $G^\ast$ to obtain realizations from the ambient posterior distribution}
		\label{fig:framework}
	\end{figure}

	GAN-Flow uses a GAN, specifically a Wasserstein GAN, to form a data-driven informative prior that can synthesize realizations of $\x$ similar to the constituents the prior dataset $\mathcal{S}$. Moreover, the generator network of the GAN becomes a map between the low-dimensional latent space and the high-dimensional ambient space. The inverse problem is solved in the low-dimensional latent space using variational Bayesian inference as the normalizing flow model acting as a pushforward operator from the prior distribution to the posterior distribution. \Cref{fig:framework} shows the three phases of the GAN-Flow framework. 
	
	\subsection{Phase A: Training a GAN-based prior}\label{subsec:GANpriors}
	GAN-Flow utilizes a Wasserstein GAN with Gradient Penalty (WGAN-GP) \cite{arjovsky2017wasserstein,gulrajani2017improved} to model the prior probability distribution $\prior$. For a WGAN-GP, the loss function $\mathcal{L}_{\mathrm{GAN}}$ is given as
	\begin{equation}
		\mathcal{L}_{\mathrm{GAN}}(\thetaa, \phii) = \Exp_{\x \sim \prior} \left[  \dm(\x, \phii) \right] - \Exp_{\z \sim \priorZ} \left[  \dm(G(\z, \thetaa), \phii) \right],
	\end{equation}
	and the min-max optimization problem in \Cref{eq:GAN-minmax} is solved under the constraint that $\dm(\z, \phii)$ lies in the space of 1-Lipschitz functions. This constraint is satisfied by enforcing a soft penalty on the gradients of the critic $\dm$ with respect to $\z$ \cite{gulrajani2017improved}. The resulting maximization problem that is solved to optimize the parameters of the discriminator is
	\begin{equation}\label{eq:wgan-gp}
		\phii^\ast = \arg\!\max_{\phii} \mathcal{L}_{\mathrm{GAN}}(\thetaa, \phii) -  \lambda  \Exp_{\tilde{\x} \sim \mathrm{p}_{\tilde{\bm{X}}}(\tilde{\x})} \left[ ( \lVert \nabla_{\tilde{\x}} \dm(\tilde{\x}, \phii) \rVert_2 -1)^2\right],
	\end{equation}
	where $\lambda$ is the gradient penalty parameter, and $\mathrm{p}_{\tilde{\bm{X}}}(\tilde{\x})$ is the uniform distribution over the straight line joining two pairs of points sampled from $\prior$ and the pushforward of $\priorZ$ by $G$. 
	The loss function in \Cref{eq:wgan-gp} minimizes the Wasserstein-1 distance between $\prior$ and the pushforward distribution of $\priorZ$ due to $\g$~\cite{patel2021gan}. 
	
	Now, let $\g^\ast$ denote the generator $\g$ with optimally chosen parameters $\thetaa^\ast$. For a perfectly trained GAN $\g^\ast$, 
	\begin{equation}\label{eq:wganprior}
		\Exp_{\x \sim \prior} \big[ m(\x) \big] = \Exp_{\z \sim \priorZ} \big[ m (\g^\ast(\z)) \big] \quad  \forall \; m \in C_b(\Ox),
	\end{equation}
	where $C_b(\cdot)$ is the space of continuously bounded functions.  
	\Cref{eq:poststat,eq:wganprior} can be combined to compute
	\begin{equation}\label{eq:modpoststats}
		\Exp_{\x \sim \posterior} \big[ \ell(\x) \big] = \Exp_{\z \sim \posteriorZ} \big[ \ell ( \g^\ast(\z)) \big] \quad \forall \; \ell \in C_b(\Ox),
	\end{equation}
	by choosing 
	\begin{equation}
		m(\z) = \frac{\ell(\z)\mathrm{p}_{\mathcal{Y}}(\hat{\y} \vert \z)}{\evidence},
	\end{equation}
	where 
	\begin{equation}\label{eq:latent_posterior}
		\posteriorZ = \frac{ \mathrm{p}_{\mathcal{Y}}(\hat{\y} \vert \z) \priorZ }{\evidence}
	\end{equation}
	is the posterior distribution of the latent variable $\z$ and $\mathrm{p}_{\mathcal{Y}}(\hat{\y} \vert \z) $ is nothing but $\likelihood$ evaluated at $\x = G^\ast(\z)$, \ie
	\begin{equation}\label{eq:yhatcondz}
		\mathrm{p}_{\mathcal{Y}}(\hat{\y} \vert \z) = \likelihood \rvert_{\x = G^\ast(\z)}.
	\end{equation}
	For additive noise models, \Cref{eq:yhatcondz} simplifies to $\mathrm{p}_{\mathcal{Y}}(\hat{\y} \vert \z) = \mathrm{p}_{\eta}(\hat{\y} - F(\g^\ast(\z)))$. Significantly, \Cref{eq:modpoststats} implies that any statistics with respect to $\x$ (respectively $\x\vert \data$) can be computed using realizations of $\z$ (respectively $\z \vert \data$). 
	
	At the end of phase A, a trained generator $\g^\ast$ is available. The GAN not only acts as a data-driven prior, but the trained generator $\g^\ast$ also serves as  an injective map that can conveniently transform realizations of the latent variable $\z$ from the latent space $\Oz$ to corresponding realizations of the ambient variable $\x$ in the ambient space $\Ox$. Additionally, $\priorZ$ is herein chosen to be an $\Nz$-variate standard normal distribution. Also note that, in most practical cases, $\Nx$ will be large but the latent dimensionality of the WGAN-GP model will be such that $\Nz \ll \Nx$. Thus, dimension reduction is achieved by using GAN priors. The next step is to solve the inverse problem in this lower-dimensional latent space. 
	
	\begin{rem}
		Choosing the latent space dimensionality $\Nz$ from $n_{\mathrm{data}}$ realizations of $\prior$ is not straightforward. This will involve careful judgment from the user. However, meta-heuristic metrics, such as the Fr\'{e}chet Inception Distance (FID) score~\cite{heusel2017gans} and Inception score~\cite{salimans2016improved}, are useful for comparing the quality of the realizations generated using GANs and real samples. These metrics can be used to evaluate the quality of the trained GAN, and suitably adjust the latent space dimensionality if performance is not satisfactory. Metrics that are more physically motivated may also be utilized in engineering applications. For instance, the consistency of miscrostructural descriptors, like the distribution of porosity in a bi-phase material~\cite{bhaduri2021efficient}, across the generated samples can be used when evaluating GANs for generating miscrostructures of heterogeneous media. We suggest choosing the smallest possible latent dimensionality that performs satisfactorily. This will also help keep the subsequent normalizing flow model relatively lightweight. 
	\end{rem}
	
	\begin{rem}
		Unless there is an influx of new prior information that must be incorporated, there is no need to retrain WGAN-GP prior. This means that Phase A, which involves the training of the WGAN-GP model, must be completed only once for a given inverse problem. Thereafter, we can reuse the trained generator for multiple inferences. 
	\end{rem}
	
	\subsection{Phase B: Inference using normalizing flows}\label{subsec:VBINF}
	
	With the latent prior density $\priorZ$ and an optimally trained generator $\g^\ast$, a normalizing flow is used to sample from the conditional posterior $\posteriorZ$. This is done by learning a bijective map $\hmm : \Oz \to \Oz$, parameterized by $\psii$, such that $\hmm_{\#}\mathrm{p}_{\mathcal{Z}}(\z; \psii) \sim \posteriorZ$. The parameters $\psii$ of the bijective map $\hmm$ are chosen by minimizing the loss function 
	\begin{equation}\label{eq:normalizingflowloss1}
		\psii^\ast = \arg \min_{\psii} d_{\mathrm{KL}} \big( \hmm_{\#}\mathrm{p}_{\mathcal{Z}}(\z; \psii) \Vert \posteriorZ \big),
	\end{equation}
	where $d_{\mathrm{KL}}(\cdot \Vert\cdot)$ is the reverse KL divergence. On simplifying \Cref{eq:normalizingflowloss1}, the loss function $	\mathcal{L}_{\mathrm{NF}}$ for training the normalizing flow takes the form:
	\begin{equation}\label{eq:normalizingflowloss}
		\mathcal{L}_{\mathrm{NF}}(\psii) = \Exp_{\z \sim \priorZ} \big[ - \log \mathrm{p}_{\mathcal{Y}}(\hat{\y} \vert \hmm(\z; \psii) ) - \log \mathrm{p}_{\mathcal{Z}}(\hmm(\z; \psii)) - \log \lvert \det \nabla_{\z} \hmm(\z;\psii) \rvert  \; \big],
	\end{equation}
	where $\mathrm{p}_{\mathcal{Y}}(\hat{\y} \vert \hmm(\z) )$ can be evaluated using \Cref{eq:yhatcondz} as 
	\begin{equation}
		\mathrm{p}_{\mathcal{Y}}(\hat{\y} \vert \hmm(\z; \psii)) = \likelihood \rvert_{\x = G^\ast(\hmm(\z))},
	\end{equation}
	and, for additive noise models, $\mathrm{p}_{\mathcal{Y}}(\hat{\y} \vert \hmm(\z; \psii)) =  \mathrm{p}_{\eta}(\hat{\y} - F(\hmm(\g^\ast(\z); \psii)))$. In \Cref{eq:normalizingflowloss},  $\nabla_{\z} \hmm(\z)$ is the Jacobian of $\hmm$. Thus, the physics model $F$ and the trained generator $G^\ast$ enters \Cref{eq:normalizingflowloss} through the log-likelihood term $\log \mathrm{p}_{\mathcal{Y}}(\hat{\y} \vert \hmm(\z; \psii))$. At the end of phase B, a trained normalizing flow map $\hmm(\cdot; \psii^\ast)$, herein denoted as $\hmm^\ast$, is available alongside the trained generator~$\g^\ast$.  
	
	\begin{rem}
		Suppose there is a change in the measurement vector $\hat{\y}$ as new measurements are available or the forward model $F$ changes, we must retrain the normalizing flow model. Perhaps one can reduce the computational burden of retraining through knowledge transfer from the previously trained normalizing flow model, for instance, by starting training from the old weights, by freezing the weights of some of the flow layers, or by simply appending new flow layers to the existing normalizing flow model. However, we do not consider such cases in this work, and the development of knowledge transfer schemes is beyond the scope of the current work.
	\end{rem}
	
	\begin{rem}
		The training of the normalizing flow model boils down to the minimization problem in \Cref{eq:normalizingflowloss1} with the loss function given by \Cref{eq:normalizingflowloss}. The minimization problem can be solved using an appropriate stochastic gradient descent algorithm. In this work, we use the Adam algorithm~\cite{kingma2014adam}. Regardless of the optimization algorithm used, training the normalizing flow using gradient descent algorithms involves the computation of the gradients of the output forward model $F$ with respect to its input. This will pose challenges when $F$ is a black-box model that is incompatible with automatic differentiation, ultimately leading to an increase in the overall computational cost; this challenge is not a bottleneck unique to GAN-Flow as advanced MCMC algorithms, like HMC, also require the gradients of $F$~\cite{patel2021solution}. One potential solution is to couple GAN-Flow with automatically differentiable surrogate models for $F$, such as neural networks~\cite{wang2022variational}, but this is beyond the study herein.
	\end{rem}
	
	\subsection{Phase C: Posterior sampling and estimation}
	
	We can use the optimally trained generator $G^\ast$ and normalizing flow map $\hmm^\ast$ to estimate $\Exp_{\x \sim \posterior} \big[ \ell(\x) \big] $ using Monte Carlo (MC) simulation. Let $\bar{\ell} $ denote the MC estimator for $\Exp_{\x \sim \posterior} \big[ \ell(\x) \big] $ given by
	\begin{equation}\label{eq:est_post_stat}
		\bar{\ell} = \frac{1}{\Ns} \sum_{i=1}^{\Ns} \ell(\g^\ast(\hmm^\ast(\z^{(i)}))),
	\end{equation}
	where $\z^{(i)}$ are independent realizations drawn from $\priorZ$. For instance, the posterior mean of $\x$ can be computed by setting $\ell(\x) = \x$. Let $\bar{\x}$ denote the MC estimator for $\Exp_{\x \sim \posterior} \big[ \x \big] $, then 
	\begin{equation}
		\bar{\x} = \frac{1}{\Ns}  \sum_{i=1}^{\Ns}  \g^\ast(\hmm^\ast(\z^{(i)})),
	\end{equation}
	where  $\z^{(i)}$ are iid realizations from $\priorZ$. Similarly, the standard deviation for the \supth{i} component of $\x$, herein denoted as $\left[\sigma_{\x}\right]_i$, can be estimated as:
	\begin{equation}
		\left\{ \sigma_{\x}\right\}_i = \sqrt{\frac{1}{\Ns-1} \sum_{j=1}^{\Ns} \big[  \left\{\g^\ast(\hmm^\ast(\z^{(j)}))\right\}_i - \{\bar{\x}\}_i \big]^2} ,    
	\end{equation}
	where $[\cdot]_i$ denotes the \supth{i} component of the vector corresponding vector. 
	
	\begin{rem}
		Neither evaluating $\hmm^\ast$ nor $\g^\ast$ requires evaluation of the underlying physics model $F$. Therefore, posterior statistics or posterior predictive quantities can be computed at almost negligible computational cost, and $\Ns$ may be set arbitrarily large. This is another advantage of the proposed GAN-Flow framework. 
	\end{rem}
	
	\subsection{Discussion of the computational cost of each phase of GAN-Flow}
	
	Phase A of GAN-Flow involves the training of the WGAN-GP model. However, GAN-Flow does not evaluate the forward model $F$ at this stage. Thus, the computational cost of Phase A will be dominated by the cost of training the WGAN-GP model. Accordingly, the computational cost of Phase A will scale with the total number of trainable parameters in the generator and the critic. We expect that more complex prior information will require expressive models with a large number of trainable parameters to develop a suitable generator and, therefore, will be more computationally expensive. Phase B, which involves training the normalizing flow model, requires repeated evaluations of the forward model $F$. Thus, the cost of evaluating the forward model $F$ will dominate the computational cost of Phase B. More complicated posterior distributions will require an expressive normalizing flow model. Expressivity can be achieved by adding more flow layers or using complex transformations, like the affine coupling transform, which will require longer training, translating to more evaluations of the forward model $F$, ultimately increasing the computational cost. Phase C neither involves training network models nor evaluating the forward model $F$. Thus, Phase C will be the least computationally demanding stage of GAN-Flow. The computational cost of this step will be dominated by the cost of evaluating the trained generator $\g^\ast$ and the trained normalizing flow model $\hmm^\ast$ and scale with the sample size $\Ns$. We envisage that Phase B will be the most computationally demanding stage of GAN-Flow in situations in physics-based applications where the underlying forward model $F$ is computationally demanding.

	\subsection{Approximation errors due to GAN-Flow}
	There are two potential sources of error in GAN-Flow when it is used in practice. The first source of error is from the WGAN-GP prior. 
	A WGAN-GP trained using \Cref{eq:wgan-gp} may only be able to approximately satisfy  \Cref{eq:wganprior} \cite{mallasto2019well} or not at all~\cite{stanczuk2021wasserstein}. The error may stem from using an approximate estimator for the Wasserstein distance in \Cref{eq:wgan-gp}, the use of MC estimates for estimating the Wasserstein distance, and failure to reach the optimal point~\cite{mallasto2019well}. However, previous research \cite{patel2021gan,patel2021solution} has shown that WGAN-GP is useful as a data-driven prior despite theoretical concerns. In practice, we monitor the loss function and terminate training when its value no longer decreases.  A second source of error stems from the normalizing flow map in situations where it is unable to induce a good approximation to the latent posterior distribution~\cite{yao2018yes}. The latent posterior distribution may not belong to the family of distributions that the bijective map is capable of inducing, which may be due to the limited fidelity (expressive power) of the normalizing flow model. The pushforward distribution may also fail to approximate the latent posterior when the reverse KL-divergence loss does not attain a value of zero, possibly due to slow convergence during training. Even if the loss function attains a value of zero, the reverse KL-divergence is known to be `mode seeking', therefore, it is possible that the pushforward distribution is unable to approximate the tails of the latent posterior. In practice, we monitor the value of the loss function $\mathcal{L}_{\mathrm{NF}}$ and stop training when it no longer decreases. However, it must be noted that, even when the pushforward distribution is poor, estimates computed using the pushforward map may continue to be useful. So, we evaluate the error in the estimated posterior statistics, such as the posterior mean and standard deviation, to determine the quality of inference. In a practical setting, we suggest the use of diagnostic tools to ascertain the quality of the pushforward distribution~\cite{yao2018yes} or stacking multiple normalizing flows with different seeds~\cite{yao2020stacking}. 
	
	\section{Results}\label{sec:results}
	In this section, we use GAN-Flow to solve three physics-based inverse problems: inference of initial conditions (\Cref{subsec:heat}), inverse Radon transform (\Cref{subsec:tomo}), and phase retrieval (\Cref{subsec:phase}). By solving these inverse problems, we demonstrate that GAN-Flow can tackle large-scale linear and non-linear inverse problems, various levels of noise, and challenging prior distributions. Our implementation also reveals that GAN-Flow is flexible in accommodating different types of WGAN-GP and normalizing flow architecture types, training methodologies, and combinations thereof. More specifically, the variety in the numerical examples we present lies in the following:
	\begin{enumerate}
		\item Forward model --- The three inverse problems we consider involve different physical phenomena. The initial condition inference problem is based on heat diffusion in a solid body, and the inverse Radon transform problem is based on an object's attenuation of penetrating waves by an object. Both aforementioned physics phenomena can be described using a linear forward model; we adapt these inverse problems from \cite{patel2019bayesian,patel2021solution}. The third problem we consider is phase retrieval, which forms the underpinnings of many modern coherent diffraction imaging methods, wherein an object is reconstructed from the magnitude of its Fourier transform. In this case, the forward model is highly nonlinear. We adapt the phase imaging problem from \cite{bohra2022bayesian}. 
		\item Prior dataset --- We consider three different prior datasets. The prior dataset for the initial condition inference and inverse Radon transform problems consists of rectangular inclusions in a zero-background and Shepp-Logan head phantoms, respectively; however, these priors are synthetic and adapted from \cite{patel2019bayesian,patel2021solution}. For the phase retrieval problem we consider a sub-sample of the publicly available NYU fastMRI~\cite{zbontar2018fastmri,knoll2020fastmri} dataset of human knee slices. 
		\item Ambient dimensionality --- The ambient dimensionality of the inverse problems we consider vary vastly. While the initial condition inference problem has a moderate ambient dimensionality of 1,024, the phase retrieval problem is a large-scale inverse problem with an ambient space of dimension 65,536, an order of magnitude beyond that of the initial condition inference problem. 
		\item WGAN-GP prior model --- For all numerical examples, we consider a WGAN-GP prior, \ie the loss function used to train the GANs is given by \Cref{eq:wgan-gp}. However, we use different architectures and/or training methodologies. The inverse Radon transform problem utilizes a simple generator and critic, consisting of fully connected and convolution layers. We use self-attention-based~\cite{zhang2019self} modules, along with convolutions, for the generator and critic for the initial condition inference problem. We found that self-attention modules help render the sharp transitions between the rectangular inclusion and the zero background. The phase retrieval problem requires still more sophisticated training to synthesize large (256$\times$256) knee slices with many fine-scale features. The WGAN-GP model for the phase retrieval problem is trained using the progressive growing of GAN methodology~\cite{karras2018progressive}. 
		\item Dimension reduction --- The WGAN-GP priors themselves lead to latent spaces of varying dimension. We work with a low-dimensional latent space for the initial condition inference problem ($\Nz = 5$), and a latent space of dimension 512 for the phase retrieval problem. On average, we can achieve approximately $\mathcal{O}(10^2)$ dimension reduction across all three inverse problems while maintaining satisfactory accuracy of the estimated statistics of the posterior distribution. 
		\item Normalizing flow model --- We use two types of flow layers. For the low to moderate dimensional latent spaces, as in the initial condition inference and inverse Radon transform problem, we employ planar flow layers. For the relatively high-dimensional latent space of the phase retrieval problem, we use affine coupling transforms to construct the normalizing flows. 
	\end{enumerate}
	\Cref{tab:problem_details} provides a summary of the inverse problems we consider, their ambient and latent space dimensionality, and the dimension reduction. 
	\begin{table}[b]
		\small
		\centering
		\caption{Summary of inverse problems we consider in this work}
		\label{tab:problem_details}
		\begin{tabular}{@{\extracolsep{2pt}} l ccc}
			\toprule[1.5pt]
			& \multicolumn{3}{c}{Inverse problem} \\
			\cline{2-4}
			& \makecell{Heat conduction\\ (\Cref{subsec:heat})} & \makecell{Radon transform\\(\Cref{subsec:tomo})} & \makecell{Phase imaging\\ (\Cref{subsec:phase})} \\
			\midrule
			Type & Linear & Linear & Non-linear  \\
			Ambient dimension $\Nx$ & 32 $\times$ 32 & 128 $\times$ 128 & 256 $\times$ 256\\
			Prior dataset & Rectangular & Shepp-Logan phantom & fastMRI~\cite{zbontar2018fastmri,knoll2020fastmri} \\
			Prior dataset size $n_{\mathrm{data}}$ & 2000 & 8000 & 29877\\
			Latent dimension $\Nz$ & 5 &  60 & 512 \\
			Dimension reduction $\Nx/\Nz$ & $\sim$200 & $\sim$273 & 128\\
			\bottomrule[1.5pt]
		\end{tabular}
	\end{table}

	We implement GAN-Flow exclusively on \texttt{PyTorch}~\cite{pytorch2019}. Where possible, we compare the posterior statistics estimated using GAN-Flow with the method outlined in \cite{patel2021solution}; herein, we refer to the latter method as GAN-HMC because it uses HMC to sample the latent posterior and estimate the posterior statistics of the ambient variable. We implement HMC within \texttt{PyTorch} using the \texttt{hamiltorch} package~\cite{cobb2021scaling}. In all cases, we specify the number of leap-frog steps to be 10, and discard 50\% of the accepted states considering burn-in. The step size is adapted during the burn-in phase so as to maintain a desired acceptance rate of 0.75.  For the initial condition inference problem, where the underlying (hidden) ambient dimensionality of the synthetic prior is small (only four), we even compare the posterior statistics estimated using GAN-Flow with MC simulation~(MCS). 
	
	For all examples, following \citet{patel2021solution}, we re-scale the prior realizations between $[-1, 1]$ before training the WGAN-GP model and use hyperbolic tangent (TanH) activation in the last layer of the generator. We invert this re-scaling operation before evaluating the likelihood function. In this way, we satisfy physical constraints such as positive values of the temperature fields for the initial condition inference problem, or positive values of material density for the inverse Radon transform problem, or positive refractive index of an object in the phase retrieval problem. 
	
	\subsection{Inferring the initial conditions in heat conduction}\label{subsec:heat}
\begin{figure}[t]
		\centering
		\begin{tikzpicture}
			\node[inner sep=0pt,outer sep=0pt] (A) at (0,0) {\includegraphics[height=1.2in]{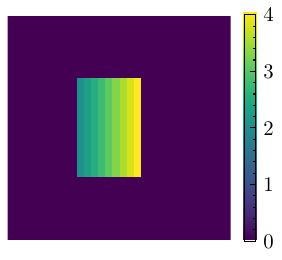}};
			\node[anchor=west, xshift=20pt, inner sep=0pt,outer sep=0pt] (B) at (A.east) {\includegraphics[height=1.2in]{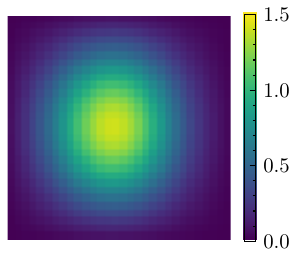}};
			\node[anchor=west, yshift=2pt, xshift=20pt, inner sep=0pt,outer sep=0pt] (C) at (B.east) {\includegraphics[height=1.16in]{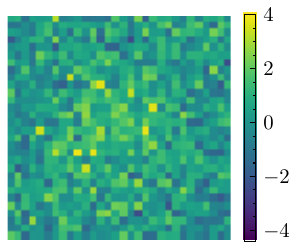}};
			\node[inner sep=0pt,anchor=north,xshift=-5pt, yshift=-5pt] at (A.north west) {(a)};
			\node[inner sep=0pt,anchor=north,xshift=-5pt, yshift=-5pt] at (B.north west) {(b)};
			\node[inner sep=0pt,anchor=north,xshift=-5pt, yshift=-5pt] at (C.north west) {(c)};
		\end{tikzpicture}
		\caption{(a) Initial and (b) final temperature fields at $t = 0$ and $1$, respectively. (c) Noisy synthetic measurements obtained after adding Gaussian white noise with unit variance to the temperature field shown in (b)}
		\label{fig:InitTemp_groundtruth}
	\end{figure}
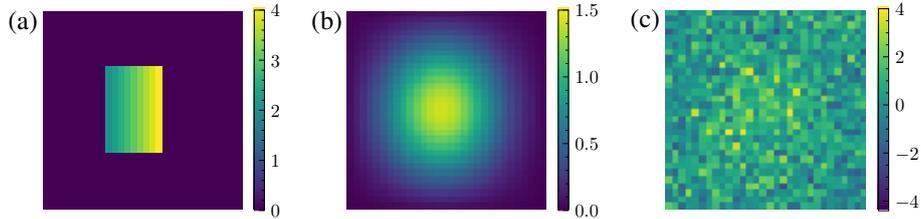

	The first problem we consider is a two-dimensional unsteady heat conduction problem where the initial condition of the temperature field $\x$ (at time $t=0$) must be inferred from a noisy measurement of the temperature field $\hat{\y}$ taken after some time (at time $t=1$). Inverse problems of this type often arise when designing thermal equipment~\cite{cannon1986inverse,cannon1998structural}. The two-dimensional time-dependent heat conduction partial differential equation over the bounded domain $\Omega$  is given as:
	\begin{equation}\label{eq:heat_pde}
		\begin{split}	
			\frac{\partial u(\sm, t)}{\partial t} - \nabla \cdot (\kappa(\sm)\nabla u(\sm, t)) &= b(\sm, t), \qquad \forall (\sm, t) \in \Omega \times (0, T) \\
			u(\sm, 0) &= m(\sm), \qquad \forall \sm \in \Omega\\
			u(\sm, t) &= 0, \qquad \forall (\sm, t) \in \partial\Omega \times (0, T)
		\end{split}
	\end{equation}
	where $T$ is the final time at which measurements are made, \ie $T=1$, and the spatial domain $\Omega$ is a square region with the length of each side being $2\pi$ units, \ie $\Omega = [0, L]\times[0, L]$ with $L = 2\pi$ units. We represent the solid on a $32 \times 32$ Cartesian grid over $\Omega$. We assume that the thermal conductivity $\kappa$ is homogeneous over $\Omega$ and equal to 0.64 units, and that $b(\sm, t) = 0$. We use the central difference scheme to discretize temperature field on the same Cartesian grid as the solid body, thus, $\Nx = \Ny = 1024$. The forward operator $F$ maps the initial temperature field $\x$ to the temperature field at time $T = 1$. We use backward-difference with a step size of 0.01 for the time-integration of \Cref{eq:heat_pde}. In this example, the inverse problem at hand is linear and it is possible to relate $\x$ and $\y$ using a linear operator, \ie $\y = \Am\x$~\cite{patel2019bayesian}.  The temperature fields at $t=0$ (initial condition) and $t=T=1$ are shown in \Cref{fig:InitTemp_groundtruth}(a) and (b), respectively. We add Gaussian white noise with unit variance to the temperature field at time $t=1$ to generate the synthetic measurements; see \Cref{fig:InitTemp_groundtruth}(c). From these noisy measurements, we want to infer the initial condition shown in \Cref{fig:InitTemp_groundtruth}(a). 
	
\begin{figure}
		\centering
		\begin{tikzpicture}
			\node[inner sep=0pt,outer sep=0pt] (A) at (0,0) {\includegraphics[height=1.8in]{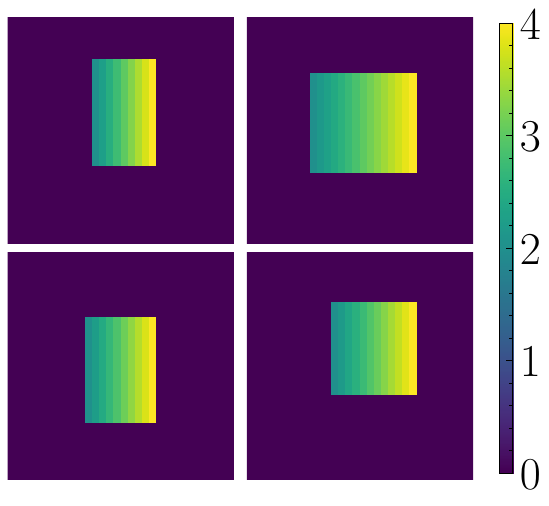}};
			\node[anchor=west, xshift=30pt, inner sep=0pt,outer sep=0pt] (B) at (A.east) {\includegraphics[height=1.8in]{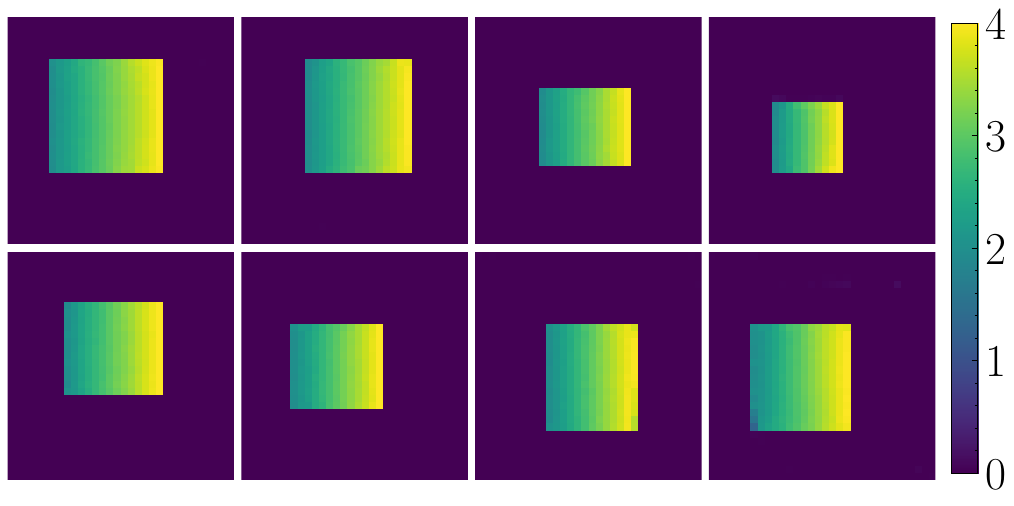}};
			\node[inner sep=0pt,anchor=north,xshift=-5pt, yshift=-5pt] at (A.north west) {(a)};
			\node[inner sep=0pt,anchor=north,xshift=-5pt, yshift=-5pt] at (B.north west) {(b)};
		\end{tikzpicture}
		\caption{(a) Realizations from the rectangular prior dataset. (b) Realizations generated from the WGAN-GP prior}
		\label{fig:rect_data_prior}
	\end{figure}

	The prior dataset consists of $n_\mathrm{data} = 2000$ initial temperature fields where the temperature is zero outside the rectangular inclusion and, within the rectangular inclusion, the temperature field varies linearly from a value of 2 units on the left edge to 4 units on the right edge. The rectangular region is generated by sampling uniformly the coordinates of the top-left and lower-right corners of the inclusion between $[0.2L, 0.4L]$ and $[0.6L, 0.8L]$, respectively. 
	We show four realizations from the prior dataset in \Cref{fig:rect_data_prior}(a). The true temperature field in \Cref{fig:InitTemp_groundtruth}(a), which we want to infer, does not belong to the prior dataset. First, we train a WGAN-GP using the prior dataset. We choose the latent space dimensionality to be 5, \ie $\Nz = 5$\footnote{We vary the latent space dimensionality $\Nz \in \{5,10,20,40,60,80,100\}$  keeping all other hyper-parameters fixed, and choose the smallest latent space dimension to yield the best estimates of the posterior mean and standard deviation;  see \Cref{appsubsec:initcond_addexp} for the results from those experiments}. Therefore, we achieve a dimension reduction of approximately 200 times in this case. For details about the WGAN-GP model and the associated hyper-parameters, see \Cref{appsubsec:gan_models} and \Cref{tab:model_details}, respectively. We show some realizations from the trained WGAN-GP prior in \Cref{fig:rect_data_prior}(b). The generated realizations are qualitatively similar to those in the prior dataset. We emphasize that training the WGAN-GP does not require any evaluation of the forward model $F$. 
	
	After we train the WGAN-GP model, we turn to training the normalizing flow model. In this example, the normalizing flow comprises 64 planar flow layers. See \Cref{tab:model_details} for more details about the hyper-parameters associated with training the normalizing flow model. Significantly, the normalizing flow model is trained for 1000 epochs with a batch size of 32, meaning a total of 32,000 evaluations of the forward model $F$.  After the normalizing flow model has been trained, we can use both the trained generator of the WGAN-GP model and the normalizing flow model to obtain as many samples from the posterior as necessary. We show the posterior mean and standard deviation of the initial temperature field estimated using GAN-Flow from a sample of size 15,000 in the left most column of \Cref{fig:initcond_poststats}. For the purposes of comparison, the posterior mean and standard deviation estimated using MCS (of sample size 10\textsuperscript{6}) and GAN-HMC are also shown in \Cref{fig:initcond_poststats}. In contrast to GAN-Flow, GAN-HMC makes 3$\times$10\superscript{5} evaluations of $F$ to yield 15,000 realizations from the posterior. \Cref{tab:initcond_rmse} tabulates the root-mean-square error of the statistics estimated using GAN-Flow and GAN-HMC, with the statistics estimated using MCS serving as the reference. From \Cref{fig:initcond_poststats}, we observe that the posterior statistics estimated using GAN-Flow and GAN-HMC compare very well with the `\emph{true}' posterior statistics estimated using MCS. Quantitatively, the posterior mean estimated using GAN-Flow is marginally better than GAN-HMC, but this improvement is achieved with greater computational efficiency (about one order of magnitude fewer evaluations of the forward model $F$). These results are promising and suggest that GAN-Flow may even be more computationally efficient than GAN-HMC.  
	\begin{table}[t!]
		\begin{minipage}[c]{0.60\linewidth}
			\centering
			\begin{tikzpicture}
				\node[inner sep=0pt,outer sep=0pt] (A) at (0,0) {\includegraphics[height=1.2in]{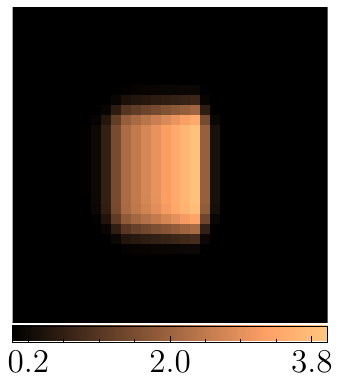}};
				\node[anchor=west, xshift=0pt, inner sep=0pt,outer sep=0pt] (B) at (A.east) {\includegraphics[height=1.2in]{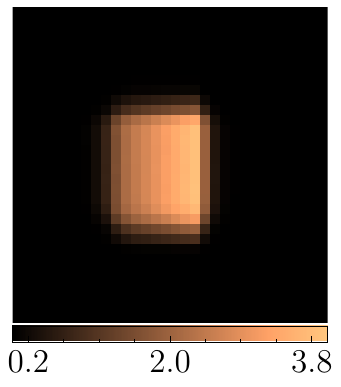}};
				\node[anchor=west, xshift=0pt, inner sep=0pt,outer sep=0pt] (C) at (B.east) {\includegraphics[height=1.2in]{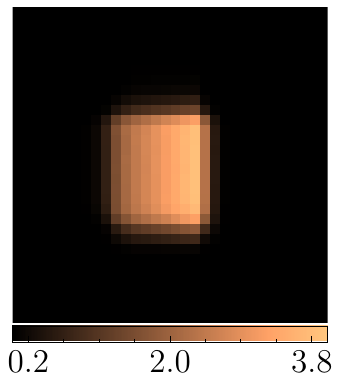}};
				
				\node[inner sep=0pt,outer sep=0pt] (A1) at (0,-3) {\includegraphics[height=1.2in]{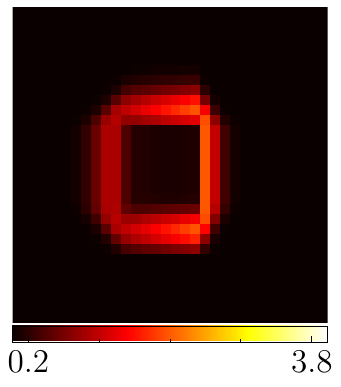}};
				\node[anchor=west, xshift=0pt, inner sep=0pt,outer sep=0pt] (B1) at (A1.east) {\includegraphics[height=1.2in]{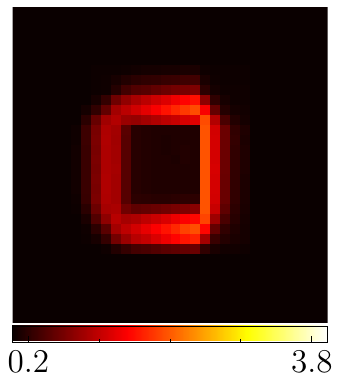}};
				\node[anchor=west, xshift=0pt, inner sep=0pt,outer sep=0pt] (C1) at (B1.east) {\includegraphics[height=1.2in]{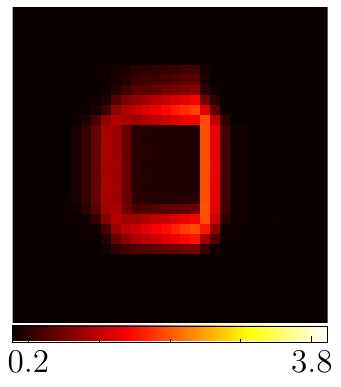}};
				
				\node[rotate=0,anchor=south,inner sep=0,outer sep=2pt] at (A.north) {MCS};
				\node[rotate=0,anchor=south,inner sep=0,outer sep=2pt] at (B.north) {GAN-Flow};
				\node[rotate=0,anchor=south,inner sep=0,outer sep=2pt] at (C.north) {GAN-HMC};
				\node[rotate=90,anchor=south,inner sep=0,outer sep=2pt,align=center] at (A.west) {Posterior\\mean};
				\node[rotate=90,anchor=south,inner sep=0,outer sep=2pt,align=center] at (A1.west) {Posterior\\std. dev.};
			\end{tikzpicture}
			\captionof{figure}{Estimated posterior mean (top row) and standard deviation (bottom row) obtained using MCS (left column), GAN-Flow (middle column) and GAN-HMC (right column) for the initial condition inference problem}
			\label{fig:initcond_poststats}
		\end{minipage}\hfill
		\begin{minipage}[c]{0.35\linewidth}
			\centering
			\caption{RMSE in the posterior statistics obtained using different inference method compared to MCS}
			\label{tab:initcond_rmse}
			\begin{tabular}{l cc}
				\toprule[1.5pt]
				\multirow{2}{*}{Method}& \multicolumn{2}{c}{Posterior statistic}\\
				\cline{2-3}
				& Mean& \makecell{Standard\\deviation} \\
				\midrule[1.5pt]
				GAN-Flow&0.034&0.048\\
				GAN-HMC&0.071&0.049\\
				\bottomrule[1.5pt]
			\end{tabular}
		\end{minipage}
	\end{table} 
	
	\subsection{Inverse Radon transform}\label{subsec:tomo}
	Next, we consider the inverse problem of reconstructing an object from its noisy sinogram. Inverse problems of this type arise in computerized tomography (CT) wherein an object is scanned from different angles using X-ray beams, and subsequently reconstructed using information about the difference in intensity before and after the beam passes through the object~\cite{feeman2015mathematics}. The forward model is given by the Radon transform. Given the material density function $\rho \in \Omega \subset \Re^2 \to \Re$, the Radon transform is defined as~\cite{feeman2015mathematics}
	\begin{equation}\label{eq:radon_transform}
		\mathcal{R}(\rho; t, \phi) = \int_{l_{t, \phi}} \!\!\!\!\rho \, \mathrm{d}l,
	\end{equation}
	where $l_{t, \phi}$ is the line that traverses through the object at a distance of $t$ from the center and an inclination of $\phi$. Therefore, given an input phantom image $\x \in \Re^{n_{\mathrm{p}} \times n_{\mathrm{p}}}$, the forward model is
	\begin{equation}\label{eq:radon_linear}
		\y = F(\x) \in \Re^{n_{\mathrm{p}} \times n_{\mathrm{p}}},
	\end{equation}
	which is a linear transformation of $\x$. In \Cref{eq:radon_linear}, 
	\begin{equation}
		\y_{i,j}= \mathcal{R}^{h}(\x; t_i, \phi_j), \;\; t_i = \frac{i}{n_{\mathrm{p}}}, \phi_j = \frac{j}{\pi} \quad \forall i,j \in \{ 1, 2, \ldots, n_{\mathrm{p}}\}, 
	\end{equation}
	and $\mathcal{R}^{h}$ is the discrete Radon transform~\cite{feeman2015mathematics}. The output $\y$ is commonly known as a sinogram. In this example, we consider input images of size 128$\times$128, \ie $n_{\mathrm{p}} = 128$. Additionally, every input image is scanned at 128 uniformly spaced angles between $0^\circ$ and $180^\circ$ with 128 detectors; thus, $\Ny = 128 \times 128$. We use the \texttt{torch-radon} package \cite{ronchetti2020torchradon} to compute Radon transforms. 
	\begin{table}[t]
		\centering
		\caption{Base parameters of the Shepp-Logan phantom. Note, $\alpha_k$ is in degrees}
		\label{tab:phantom_parameters}
		\begin{tabular}{c r r r r r r}
			\toprule[1.5pt]
			$k$ & $r_k$ & $s_k$ & $a_k$& $b_k$& $\alpha_k$& $\rho_k$\\
			\midrule
			1 & 0.0 	& 0.0		&0.69 &0.92 &0 &1.0 \\
			2 & 0.0 	& $-$0.0184	&0.6624 &0.874 &0 &$-$0.8 \\
			3 & 0.22 	& 0.0		&0.11 &0.31 &$-$18 &$-$0.2 \\
			4 & $-$0.22 & 0.0		&0.16 &0.41 &$-$18 &$-$0.2 \\
			5 & 0.0		& 0.35		&0.21 &0.25 &0 &0.1 \\
			6 & 0.0		& 0.1		&0.046 &0.026 &0 &0.1 \\
			7 & 0.0		& $-$0.1	&0.046 &0.046 &0 &0.1 \\
			8 & $-$0.08	& $-$0.605	&0.046 &0.023 &0 &0.1 \\
			9 & 0.0		& $-$0.606	&0.023 &0.023 &0 &0.1 \\
			10 & 0.06	& $-$0.605	&0.023 &0.046 &0 &0.1 \\
			\bottomrule[1.5pt]
		\end{tabular}
	\end{table}
	The prior dataset for this example consists of Shepp-Logan head phantoms~\cite{toft1996radon}. Every phantom consists of ten ellipses, where each ellipse has constant density. Let the \supth{k} ellipse $E_k$ be centered at $(r_k, s_k)$, with semi-axis lengths $a_k$ and $b_k$, angle of inclination $\alpha_k$ (in degrees) and density $\rho_k$. The density of the phantom at any coordinate $(r,s)$ is
	\begin{equation}
		\rho(r,s) = \Sigma_{k=1}^{10} C_k(r,s), \quad \text{ where }  C_k(r,s) = \begin{dcases}
			\rho_k & \text{ if } (r,s) \in E_k \\
			0 & \text{ otherwise }
		\end{dcases}
	\end{equation}
	\Cref{tab:phantom_parameters} provides details of the nominal values of the base parameters of the ellipses, which we adapt from~\citet{toft1996radon}. Following \citet{patel2021solution}, we generate new phantoms by perturbing the base parameters of every ellipse $\tilde{E}_k$ as follows:
	\begin{equation}
		\begin{split}
			&\tilde{r}_k = r_k + 0.005\xi_{k,1}, \;\; \tilde{s}_k = s_k + 0.005\xi_{k,2}, \;\; \tilde{a}_k = a_k + 0.005\xi_{k,3}, \\
			&\tilde{b}_k = b_k + 0.005\xi_{k,4}, \;\; \tilde{\alpha}_k = \alpha_k + 2.5\xi_{k,5}, \;\; \tilde{\rho}_k = \rho_k + 0.0005\xi_{k,6}
		\end{split}
	\end{equation}
	where $\left\{ \left\{ \xi_{k,i} \right\}_{i=1}^{i=6} \right\}_{k=1}^{k=10}$ are uniform random variables in $[-1,1]$. Now, the density of the perturbed phantom $\tilde{\rho}$ is
	\begin{equation}\label{eq:phantom_perturbed}
		\tilde{\rho}(r,s) = \max\bigg(0, \min\Big( 1, \Sigma_{k=1}^{10} \tilde{C}_k(r,s)\Big)\bigg), \quad \text{ where }  \tilde{C}_k(r,s) = \begin{dcases}
			\tilde{\rho}_k & \text{ if } (r,s) \in \tilde{E}_k \\
			0 & \text{ otherwise }
		\end{dcases}
	\end{equation}
	ensures that the material density $\tilde{\rho}$ at any point is bounded within 0 (air cavity) and 1 (bone). We obtain $n_{\mathrm{data}} = 8000$ discrete phantom images by evaluating \Cref{eq:phantom_perturbed} on a grid of size $128 \times 128$. The resulting image is further subject to a transformation that translates it by $n_{\mathrm{h}}$ and $n_{\mathrm{v}}$ pixels in the horizontal and vertical direction, respectively, and rotates it by an angle $\beta$. We assume that $n_{\mathrm{h}}$ and $n_{\mathrm{v}}$ take integer values uniformly between $-8$ and $8$, \ie 
	$n_{\mathrm{h}}, n_{\mathrm{v}} \sim \mathcal{U}\left\{ -8, -7, \ldots, 7, 8 \right\}$,  
	while the random variable $\beta\in \mathcal{U}(-20^\circ, 20^\circ)$.
	
	\begin{figure}[t]
		\centering
		\begin{tikzpicture}
			\node[inner sep=0pt,outer sep=0pt] (A) at (0,0) {\includegraphics[height=1.8in]{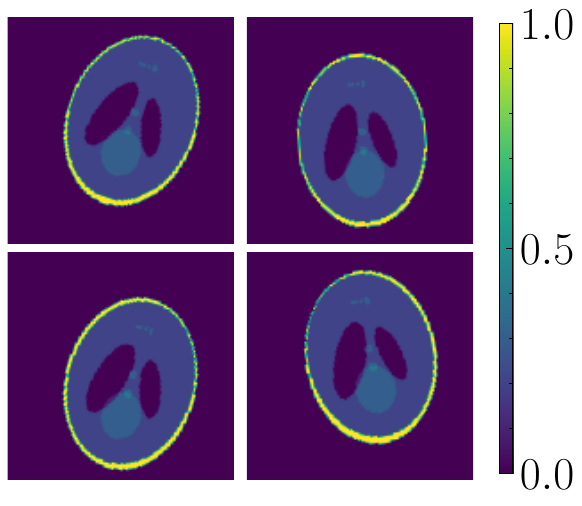}};
			\node[anchor=west, xshift=30pt, inner sep=0pt,outer sep=0pt] (B) at (A.east) {\includegraphics[height=1.8in]{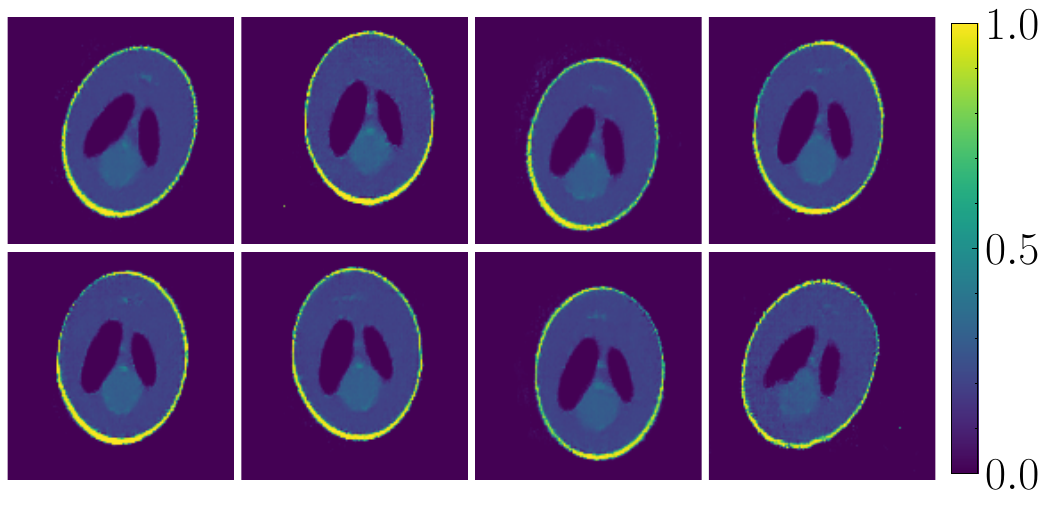}};
			\node[inner sep=0pt,anchor=north,xshift=-5pt, yshift=-5pt] at (A.north west) {(a)};
			\node[inner sep=0pt,anchor=north,xshift=-5pt, yshift=-5pt] at (B.north west) {(b)};
		\end{tikzpicture}
		\caption{(a) Different phantoms from the prior dataset (b) Phantoms generated from the WGAN-GP prior}
		\label{fig:tomo_prior}
	\end{figure}
	\begin{figure}[t]
		\centering
		\begin{tikzpicture}
			\node[inner sep=0pt,outer sep=0pt] (A) at (0,0) {\includegraphics[height=1.1in]{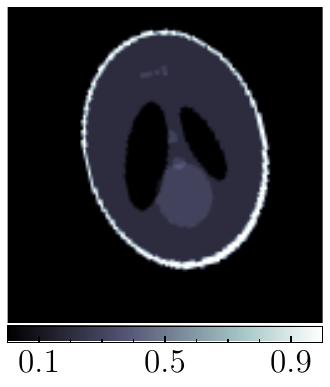}};
			\node[anchor=west, xshift=20pt, inner sep=0pt,outer sep=0pt] (B) at (A.east) {\includegraphics[height=1.1in]{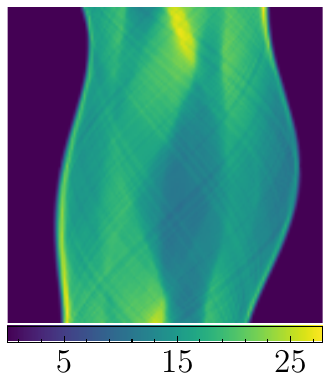}};
			\node[anchor=west, xshift=20pt, inner sep=0pt,outer sep=0pt] (C) at (B.east) {\includegraphics[height=1.1in]{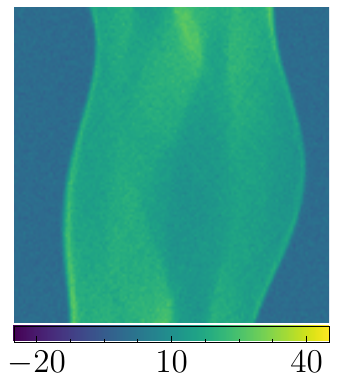}};
			\node[anchor=west, xshift=20pt, inner sep=0pt,outer sep=0pt] (D) at (C.east) {\includegraphics[height=1.1in]{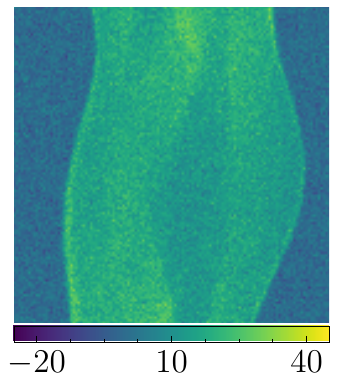}};
			\node[anchor=west, xshift=20pt, inner sep=0pt,outer sep=0pt] (E) at (D.east) {\includegraphics[height=1.1in]{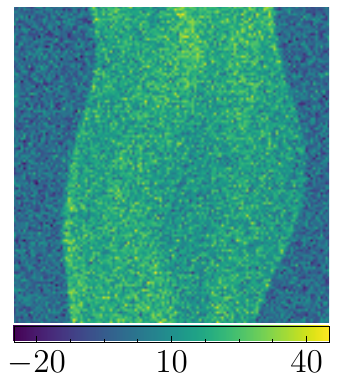}};
			
			\node[inner sep=0pt,anchor=north,xshift=-5pt, yshift=-5pt] at (A.north west) {(a)};
			\node[inner sep=0pt,anchor=north,xshift=-5pt, yshift=-5pt] at (B.north west) {(b)};
			\node[inner sep=0pt,anchor=north,xshift=-5pt, yshift=-5pt] at (C.north west) {(c)};
			\node[inner sep=0pt,anchor=north,xshift=-5pt, yshift=-5pt] at (D.north west) {(d)};
			\node[inner sep=0pt,anchor=north,xshift=-5pt, yshift=-5pt] at (E.north west) {(e)};
		\end{tikzpicture}
		\caption{(a) Test phantom and (b) corresponding noise-free sinogram. Noisy sinograms after adding zero-mean Gaussian noise with variance (c) $\sigma_\eta^2 = 1$ (d) $\sigma_\eta^2 = 10$ and (e) $\sigma_\eta^2 = 50$}
		\label{fig:tomo_groundtruth}
	\end{figure}
	
	We show four realizations from the prior dataset in \Cref{fig:tomo_prior}(a). We use another realization, not part of the prior dataset and shown in \Cref{fig:tomo_groundtruth}, to generate the synthetic measurements for this example. We simulate noisy sinogram data by adding zero-mean Gaussian noise with variance $\sigma_\eta^2$ to the noise-free sinogram. We vary the variance of the measurement noise $\sigma_\eta^2 \in \{1, 10, 50\}$ to test the robustness of GAN-Flow to varying levels of noise in the measurement. The noise characteristics of CT data is Gaussian when the photon counts are large~\cite{wang2008experimental}. A Gaussian noise model is useful even in the small photon count regime~\cite{wang2008experimental}. In this work, we limit our exposition to Gaussian noise. We remark that we can accommodate any noise model by suitably modifying the likelihood term in \Cref{eq:normalizingflowloss}. 
	
	\begin{figure}[t]
		\centering
		\begin{tikzpicture}
			\node[inner sep=0pt,outer sep=0pt] (A) at (0,0) {\includegraphics[height=1.2in]{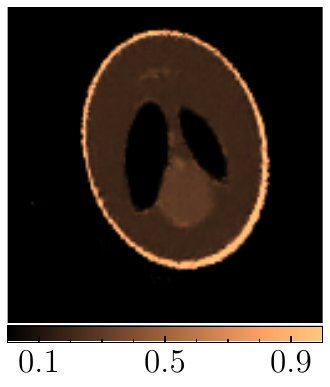}};
			\node[inner sep=0pt,outer sep=0pt,anchor=north] (B) at (A.south) {\includegraphics[height=1.2in]{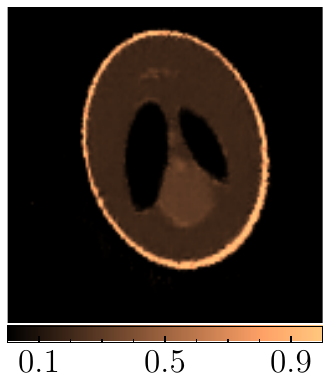}};
			\node[inner sep=0pt,outer sep=0pt,anchor=north] (C) at (B.south) {\includegraphics[height=1.2in]{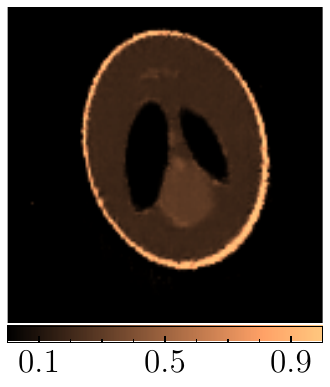}};
			
			\node[inner sep=0pt,outer sep=0pt,anchor=west] (A1) at (A.east) {\includegraphics[height=1.2in]{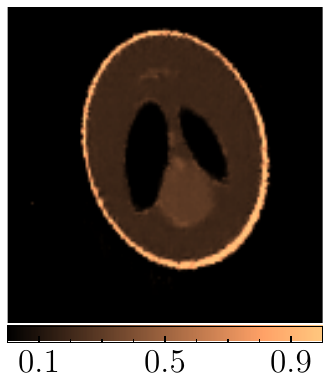}};
			\node[inner sep=0pt,outer sep=0pt,anchor=north] (B1) at (A1.south) {\includegraphics[height=1.2in]{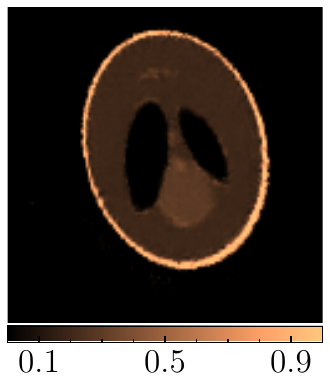}};
			\node[inner sep=0pt,outer sep=0pt,anchor=north] (C1) at (B1.south) {\includegraphics[height=1.2in]{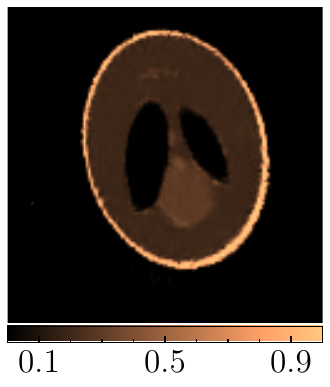}};
			
			\node[inner sep=0pt,outer sep=0pt,anchor=west, xshift=45pt] (A2) at (A1.east) {\includegraphics[height=1.2in]{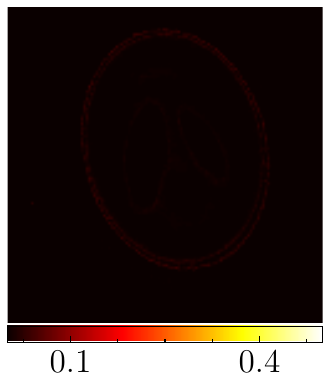}};
			\node[inner sep=0pt,outer sep=0pt,anchor=north] (B2) at (A2.south) {\includegraphics[height=1.2in]{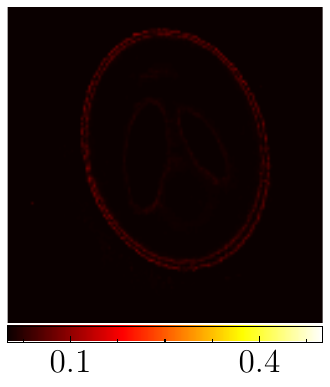}};
			\node[inner sep=0pt,outer sep=0pt,anchor=north] (C2) at (B2.south) {\includegraphics[height=1.2in]{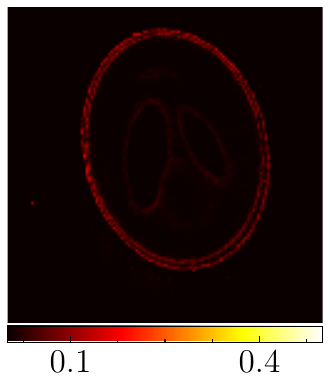}};
			
			\node[inner sep=0pt,outer sep=0pt,anchor=west] (A3) at (A2.east) {\includegraphics[height=1.2in]{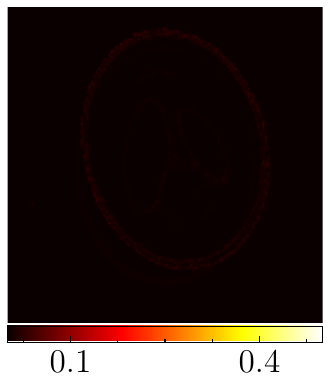}};
			\node[inner sep=0pt,outer sep=0pt,anchor=north] (B3) at (A3.south) {\includegraphics[height=1.2in]{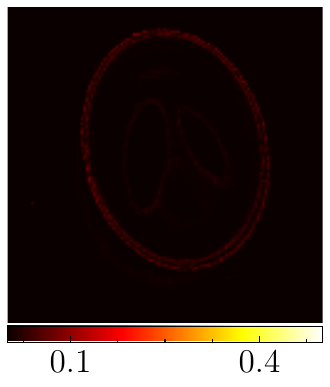}};
			\node[inner sep=0pt,outer sep=0pt,anchor=north] (C3) at (B3.south) {\includegraphics[height=1.2in]{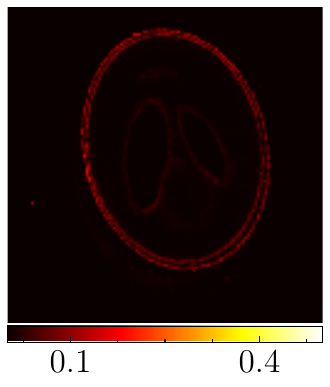}};

			\node[rotate=0,anchor=south,inner sep=0,outer sep=2pt] at (A.north) {GAN-Flow};
			\node[rotate=0,anchor=south,inner sep=0,outer sep=2pt] at (A1.north) {GAN-HMC};
			\node[rotate=0,anchor=south,inner sep=0,outer sep=2pt] at (A2.north) {GAN-Flow};
			\node[rotate=0,anchor=south,inner sep=0,outer sep=2pt] at (A3.north) {GAN-HMC};
			
			\node[rotate=0,anchor=north,inner sep=0,outer sep=2pt,align=center, font=\large,yshift=-5pt,xshift=40pt] at (C.south) {Posterior mean};
			\node[rotate=0,anchor=north,inner sep=0,outer sep=2pt,align=center, font=\large,yshift=-5pt,xshift=40pt] at (C2.south) {Posterior std. dev.};
			
			\node[rotate=90,anchor=south,inner sep=0,outer sep=2pt,align=center, font=\large,xshift=2pt,yshift=5pt] at (A.west) {$\sigma_\eta^2 = 1$};
			\node[rotate=90,anchor=south,inner sep=0,outer sep=2pt,align=center, font=\large,xshift=2pt,yshift=5pt] at (B.west) {$\sigma_\eta^2 = 10$};
			\node[rotate=90,anchor=south,inner sep=0,outer sep=2pt,align=center, font=\large,xshift=2pt,yshift=5pt] at (C.west) {$\sigma_\eta^2 = 50$};		
			\node[rotate=90,anchor=south,inner sep=0,outer sep=2pt,align=center, font=\large,xshift=2pt,yshift=5pt] at (A2.west) {$\sigma_\eta^2 = 1$};
			\node[rotate=90,anchor=south,inner sep=0,outer sep=2pt,align=center, font=\large,xshift=2pt,yshift=5pt] at (B2.west) {$\sigma_\eta^2 = 10$};
			\node[rotate=90,anchor=south,inner sep=0,outer sep=2pt,align=center, font=\large,xshift=2pt,yshift=5pt] at (C2.west) {$\sigma_\eta^2 = 50$};		
			
		\end{tikzpicture}
		\caption{Estimated posterior mean (left) and standard deviation (right) obtained using GAN-Flow (left column) and GAN-HMC (right column) for the inverse Radon transform problem at various levels of measurement noise}
		\label{fig:tomo_poststats}
	\end{figure}
	As in the previous example, we first train a WGAN-GP with latent dimensionality $\Nz = 60$ to approximate the prior distribution\footnote{Like the previous example, we vary $\Nz \in \{5,10,20,40,60,80,100\}$ and choose $\Nz = 60$ since the RMSE between the corresponding posterior mean and the test phantom is is either smallest or close to being the smallest; see \Cref{appsubsec:tomo_addexp}}. We provide details of the generator and critic architectures used in this study in \Cref{appsubsec:gan_models}, while \Cref{tab:model_details} provides details of other training hyper-parameters. \Cref{fig:tomo_prior}(b) shows some realizations from the trained WGAN-GP prior. Next, we train a normalizing model that has 256 planar flow layers. We train the normalizing flow model for 15,000 epochs with a batch size of 32; we list other hyper-parameters associated with the training in \Cref{tab:model_details}. Thus, training the normalizing flow model requires 4.8$\times$10\textsuperscript{5} forward model evaluations. After this we obtain 30,000 realizations from the posterior distribution and use them to estimate the posterior mean and standard deviation. \Cref{fig:tomo_poststats} shows the posterior statistics estimated using GAN-Flow and GAN-HMC. For comparison, we also obtain a sample of size 60,000 from the posterior distribution using HMC, discard the first 30,000 realizations considering burn-in, and then estimate the posterior statistics. 
	For this example,  we run HMC with 10 steps and an initial step size of 0.01. With this setting, sampling from the latent posterior using HMC requires 6$\times$10\textsuperscript{5} forward model evaluations. The posterior statistics estimated using GAN-Flow and GAN-HMC are qualitatively similar and shows elevated uncertainty around the edges of the phantom. The uncertainty increases as the noise in the measurement increases, which is also expected. We compute the RMSE and structural similarity index metric (SSIM)~\cite{wang2004image} between the posterior mean and the test phantom for both GAN-Flow and GAN-HMC and report those values in \Cref{tab:tomo_rmse}. Quantitatively, both GAN-Flow and GAN-HMC provide similar reconstructions of the test phantom, which is consistent with \Cref{fig:tomo_poststats}. The results confirm that GAN-Flow is robust with respect to the level of measurement noise. 
	\begin{table}[H]
		\centering
		\caption{RMSE and SSIM of the posterior mean reconstruction of the test phantom at different levels of measurement noise}
		\label{tab:tomo_rmse}
		\begin{tabular}{@{\extracolsep{2pt}}l ccc ccc}
			\toprule[1.5pt]
			\multirow{2}{*}{Method}& \multicolumn{3}{c}{RMSE} & \multicolumn{3}{c}{SSIM}\\
			\cline{2-4}\cline{5-7}
			\Tstrut
			& $\sigma_\eta^2 = 1$& $\sigma_\eta^2 = 10$ & $\sigma_\eta^2 = 50$ & $\sigma_\eta^2 = 1$& $\sigma_\eta^2 = 10$ & $\sigma_\eta^2 = 50$ \\
			\midrule[1.5pt]
			GAN-Flow& 0.041& 0.042& 0.045& 0.968& 0.964& 0.963\\
			GAN-HMC& 0.041& 0.043& 0.045& 0.968& 0.964& 0.962\\
			\bottomrule[1.5pt]
		\end{tabular}
	\end{table}
	
	\subsection{Phase retrieval}\label{subsec:phase}
	The final application we consider concerns phase retrieval, which involves the recovery of an object from the magnitude of its Fourier transform~\cite{klibanov1995phase,rosenblatt1984phase}. Phase retrieval inverse problems are ubiquitous in many areas of science and engineering~\cite{shechtman2015phase,miao2011coherent,maleki1993phase,millane1990phase,fienup1993hubble}. More specifically, we consider the phase retrieval problem of recovering an object from sparse measurements of the magnitude of its Fourier transform. We undersample the measurements to simulate accelerated measurement acquisition paradigms. The forward model for the phase retrieval problem we consider is given by:
	\begin{equation}\label{eq:phase_forward}
		\y = \lvert \Mm \Fm \x \rvert+ \bm{\eta},
	\end{equation}
	where $\x \in \Re^{n_{\mathrm{p}} \times n_{\mathrm{p}}}$ is the object of interest discretized as an image of $n_{\mathrm{p}} \times n_{\mathrm{p}}$ pixels, $\Fm$ is the two-dimensional discrete Fourier transform (DFT),  $\lvert \cdot \rvert$ computes the magnitude component wise, $\Mm$ is a binary mask that undersamples the Fourier magnitude measurements, and $\bm{\eta}$ is the measurement noise. In vector form, the effective dimensionality $\Ny$ of $\y$ depends on the undersampling ratio $r$ (also known as acceleration factor~\cite{zbontar2018fastmri}), \ie $\Ny = \Nx/r$. 
	
\begin{figure}
		\centering
		\begin{tikzpicture}
			\node[inner sep=0pt,outer sep=0pt] (A) at (0,0) {\includegraphics[height=1.8in]{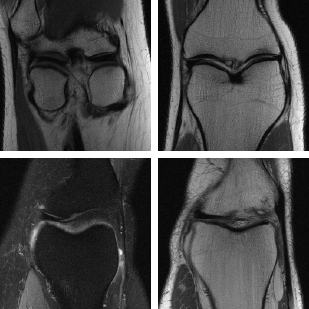}};
			\node[anchor=west, xshift=30pt, inner sep=0pt,outer sep=0pt] (B) at (A.east) {\includegraphics[height=1.8in]{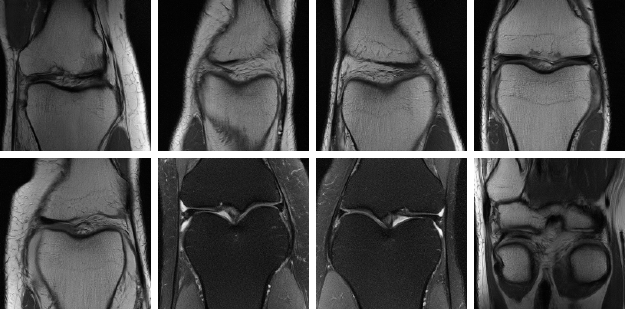}};
			\node[inner sep=0pt,anchor=north,xshift=-8pt, yshift=0pt] at (A.north west) {(a)};
			\node[inner sep=0pt,anchor=north,xshift=-8pt, yshift=0pt] at (B.north west) {(b)};
		\end{tikzpicture}
		\caption{(a) Different knee slices from the prior dataset (b) Knee slices generated from the WGAN-GP prior}
		\label{fig:phase_prior}
	\end{figure}

	For this example, the prior dataset comprises of a subsample of the single coil knee scans from the publicly available NYU fastMRI training dataset~\cite{zbontar2018fastmri,knoll2020fastmri}. Similar to~\citet{kelkar2021compressible}, we prepare the prior dataset in the following way. The training dataset contains a total of 973 volumes and 34,742 slices. Each slice corresponds to an \emph{emulated} single coil complex-valued Fourier space ($k$-space) MRI measurement; the single coil data is emulated by linearly combining multi-coil $k$-space data~\cite{zbontar2018fastmri}. For every slice, the fastMRI initiative also provides a corresponding slice of the knee for that volume computed from the emulated single coil measurement using the root-sum-of-squares method. We discard the first five reconstructed knee slices of every volume, center crop the rest into images of size $256 \times 256$ and then randomly divide them up into training and test sets. In total, the training and test set of the WGAN-GP contains 29,877 and 6,140 knee slices, respectively. The aforementioned training set is the prior dataset for this example and it contains $n_{\mathrm{data}} = 29,877$ knee slices. Moreover, the ambient dimensionality $\Nx = 256 \times 256$ for this problem. \Cref{fig:phase_prior}(a) shows four typical knee slices from the prior dataset. We emphasize that, although we use knee slices from reconstructed MRI data, the forward model is nonlinear and given by \Cref{eq:phase_forward}. 
	
	We use three realizations from the test set as the test case for this problem. The test cases are shown in the top row of \Cref{fig:phase_testsamples}. Right below them, we plot the natural logarithm of the corresponding noise free $k$-space magnitude data to which we subsequently add zero-mean Gaussian noise with standard deviation equal to 0.04\% of the zero-frequency amplitude of the two-dimensional DFT~\cite{sun2021deep}. We also consider two types of Cartesian undersampling masks to reflect realistic scenarios where an object must be reconstructed from sparse measurements. Specifically, we consider two masks that yield four-fold and eight-fold accelerations. Following \citet{zbontar2018fastmri}, the undersampling masks include 8\% and 4\% of the central region of the $k$-space when the acceleration factor $r=4$ and $8$, respectively. The remaining $k$-space lines are uniformly sampled with probability such that the desired acceleration can be achieved. As common in practice, we omit $k$-space magnitude measurements in the phase direction, \ie the undersampling masks consist of vertical bands. \Cref{fig:phase_masks} shows the two masks considered in this example.
	
	\begin{table}[t!]
		\begin{minipage}[b]{0.65\linewidth}
			\centering
			\begin{tikzpicture}
				%
				%
				
				\node[inner sep=0pt, outer sep=0pt] (A) at (0,0) {\includegraphics[width=1.2in]{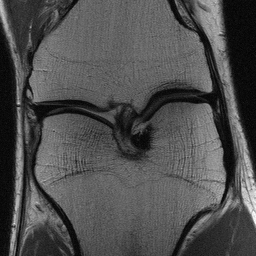}};
				\node[inner sep=0pt, outer sep=0pt] (B) at (3.5,0) {\includegraphics[width=1.2in]{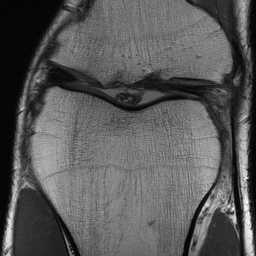}};
				\node[inner sep=0pt, outer sep=0pt] (C) at (7,0) {\includegraphics[width=1.2in]{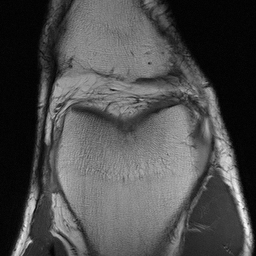}};
				
				\node[inner sep=0pt, outer sep=0pt,anchor=north,yshift=-5pt] (A1) at (A.south) {\includegraphics[width=1.26in]{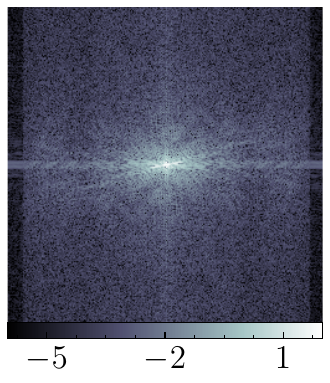}};
				\node[inner sep=0pt, outer sep=0pt,anchor=north,yshift=-5pt] (B1) at (B.south) {\includegraphics[width=1.26in]{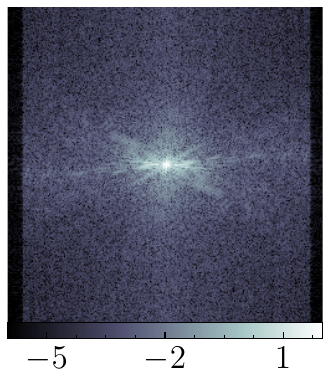}};
				\node[inner sep=0pt, outer sep=0pt,anchor=north,yshift=-5pt] (C1) at (C.south) {\includegraphics[width=1.26in]{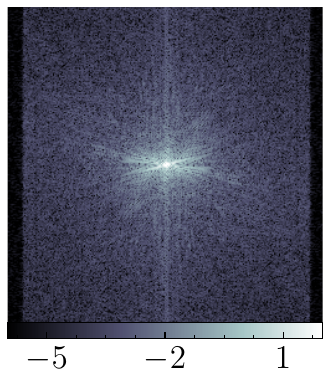}};
				
				\node[rotate=90,anchor=south,yshift=5pt] at (A.west) {Test slice};
				\node[rotate=90,anchor=south,yshift=5pt,xshift=5pt] at (A1.west) {$\log$ $k$-space mag.};

			\end{tikzpicture}
			\captionof{figure}{Test slices used for phase retrieval (top row) and corresponding noise free logarithm of Fourier ($k$-space) magnitudes (bottom row). We apply the masks, shown in \Cref{fig:phase_masks}, to the noisy Fourier magnitudes to generate the synthetic measurements for this example}
			\label{fig:phase_testsamples}
		\end{minipage}
		\hfill
		\begin{minipage}[b]{0.25\linewidth}
			\centering
			\begin{tikzpicture}
				\node[inner sep=0pt, outer sep=0pt] (A) at (0,0) {\includegraphics[width=1.2in]{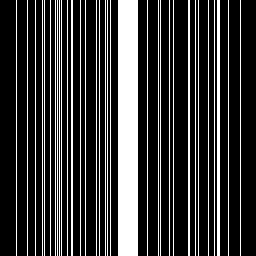}};
				\node[inner sep=0pt, outer sep=0pt,anchor=north,yshift=-7pt] (B) at (A.south) {\includegraphics[width=1.2in]{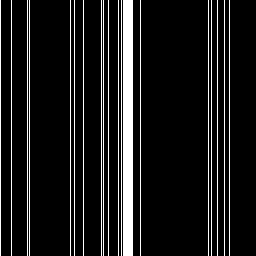}};
				\draw[white, very thick] (0,-5) rectangle (1,-5.35);
				\node[rotate=90,anchor=south,yshift=5pt] at (A.west) {4$\times$ accel.};
				\node[rotate=90,anchor=south,yshift=5pt] at (B.west) {8$\times$ accel.};
			\end{tikzpicture}
			\captionof{figure}{Cartesian undersampling masks with four- and eight-fold  accelerations}
			\label{fig:phase_masks}
		\end{minipage}
	\end{table}
	In this example, the WGAN-GP prior is trained using the progressive growing of GAN (ProGAN) method~\cite{karras2018progressive}. Not only does the ProGAN training method stabilize the training of GANs designed to synthesize large images, but it also makes the training more efficient. In the ProGAN training methodology, learning commences from a coarse scale wherein the generator learns to synthesize, and the critic learns to discriminate, low resolution images, say of size $4\times4$. Over stages of increasing resolution, going from $4\times4$ to $8\times8$ and ultimately to $256\times256$, new layers are added to the generator and the critic, as the generator learns to synthesize, and the critic learns to discriminate, finer scale details. We adopt the implementation of ProGANs from~\cite{progantorch} and choose the latent space dimensionality $\Nz=512$ following previous works~\cite{kelkar2021compressible,karras2018progressive}. Further details about the WGAN-GP model, ProGAN training methodology, and associated training hyper-parameters may be found in \ref{appsubsec:models} and \Cref{tab:model_details}. Samples from the trained WGAN-GP prior are shown in \Cref{fig:phase_prior}(b). We note that the WGAN-GP model is frozen for all subsequent steps. 
	
\begin{figure}[t]
		\centering
		\begin{tikzpicture}
			\node[inner sep=0pt, outer sep=0pt,anchor=north] (X) at (0, 0) {\includegraphics[width=1.2in]{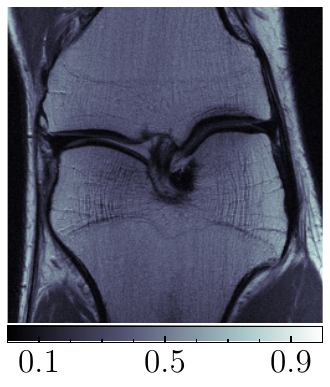}};
			\node[inner sep=0pt, outer sep=0pt,anchor=north] (A) at (3.5, 0) {\includegraphics[width=1.2in]{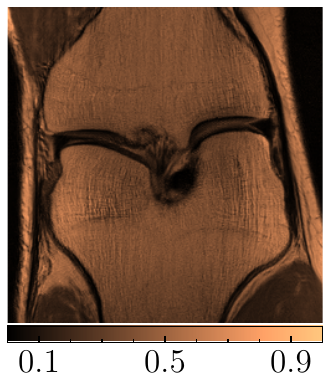}};
			\node[inner sep=0pt, outer sep=0pt,anchor=north] (B) at (7, 0) {\includegraphics[width=1.2in]{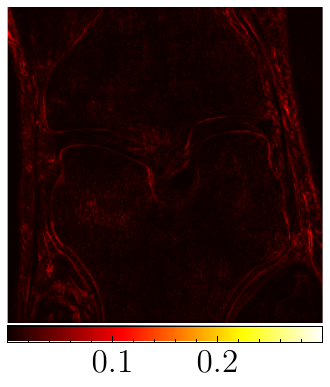}};
			\node[inner sep=0pt, outer sep=0pt,anchor=north] (C) at (10.5, 0) {\includegraphics[width=1.2in]{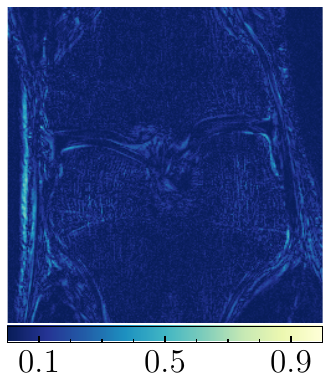}};
			\node[inner sep=0pt, outer sep=0pt,anchor=north,yshift=-5pt] (X1) at (X.south) {\includegraphics[width=1.2in]{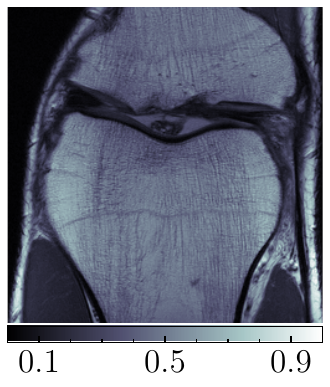}};
			\node[inner sep=0pt, outer sep=0pt,anchor=north,yshift=-5pt] (A1) at (A.south) {\includegraphics[width=1.2in]{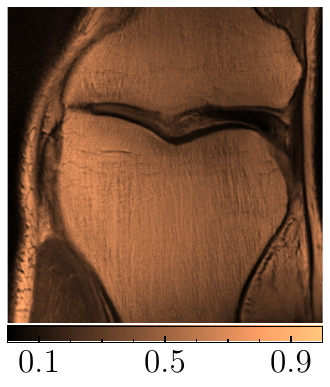}};
			\node[inner sep=0pt, outer sep=0pt,anchor=north,yshift=-5pt] (B1) at (B.south) {\includegraphics[width=1.2in]{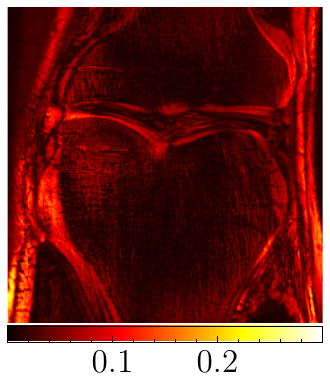}};
			\node[inner sep=0pt, outer sep=0pt,anchor=north,yshift=-5pt] (C1) at (C.south) {\includegraphics[width=1.2in]{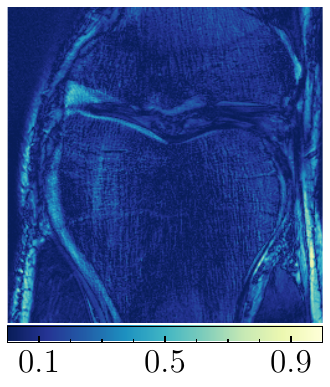}};		
			\node[inner sep=0pt, outer sep=0pt,anchor=north,yshift=-5pt] (X2) at (X1.south) {\includegraphics[width=1.2in]{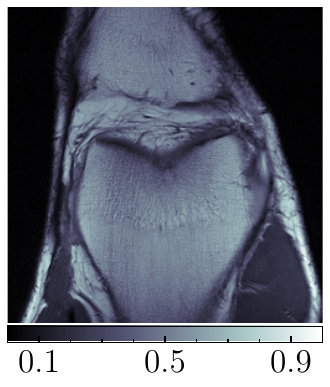}};
			\node[inner sep=0pt, outer sep=0pt,anchor=north,yshift=-5pt] (A2) at (A1.south) {\includegraphics[width=1.2in]{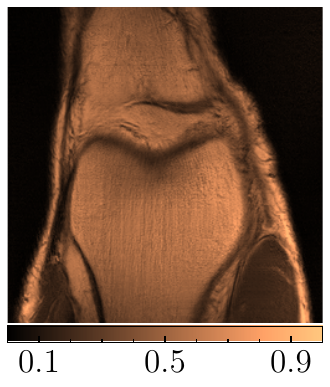}};
			\node[inner sep=0pt, outer sep=0pt,anchor=north,yshift=-5pt] (B2) at (B1.south) {\includegraphics[width=1.2in]{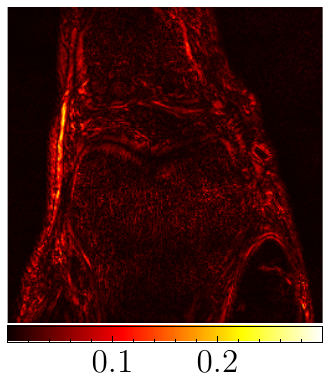}};
			\node[inner sep=0pt, outer sep=0pt,anchor=north,yshift=-5pt] (C2) at (C1.south) {\includegraphics[width=1.2in]{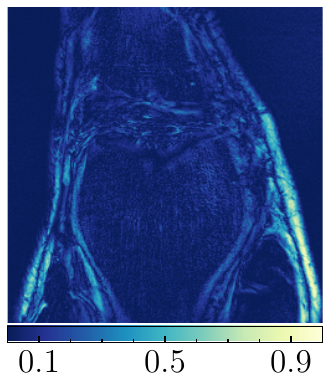}};
			\node[anchor=south] at (X.north) {Ground truth};
			\node[anchor=south] at (A.north) {Post. mean};
			\node[anchor=south] at (B.north) {Post. std. dev.};
			\node[anchor=south] at (C.north) {Abs. error};
			\node[anchor=south,rotate=90,yshift=2pt] at (X.west) {Test slice 1};
			\node[anchor=south,rotate=90,yshift=2pt] at (X1.west) {Test slice 2};
			\node[anchor=south,rotate=90,yshift=2pt] at (X2.west) {Test slice 3};
		\end{tikzpicture}
		\caption{Posterior pixel-wise standard mean (second column) and posterior pixel-wise standard deviation (third column) for various test cases when acceleration factor $r=4$. The first column shows the ground truth slices for reference. The last column shows the absolute error between the posterior mean and the ground truth}
		\label{fig:phase_recon_accel_4}	
	\end{figure}

\begin{figure}[t]
		\centering
		\begin{tikzpicture}
			\node[inner sep=0pt, outer sep=0pt,anchor=north] (X) at (0, 0) {\includegraphics[width=1.2in]{scan_2.pdf}};
			\node[inner sep=0pt, outer sep=0pt,anchor=north] (A) at (3.5, 0) {\includegraphics[width=1.2in]{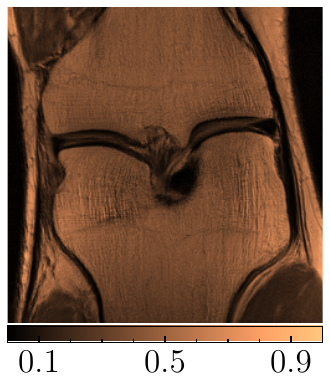}};
			\node[inner sep=0pt, outer sep=0pt,anchor=north] (B) at (7, 0) {\includegraphics[width=1.2in]{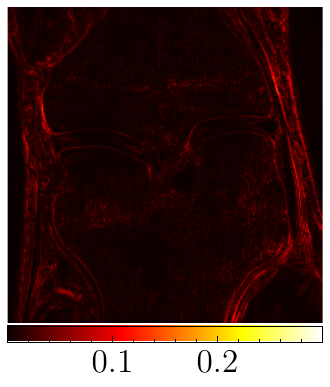}};
			\node[inner sep=0pt, outer sep=0pt,anchor=north] (C) at (10.5, 0) {\includegraphics[width=1.2in]{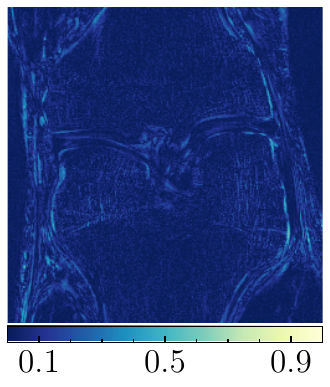}};
			\node[inner sep=0pt, outer sep=0pt,anchor=north,yshift=-5pt] (X1) at (X.south) {\includegraphics[width=1.2in]{scan_3.pdf}};
			\node[inner sep=0pt, outer sep=0pt,anchor=north,yshift=-5pt] (A1) at (A.south) {\includegraphics[width=1.2in]{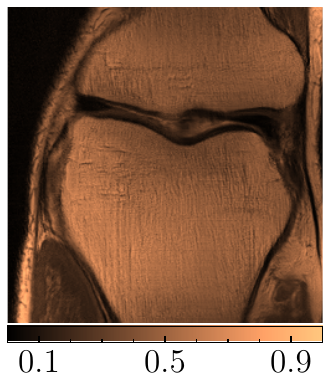}};
			\node[inner sep=0pt, outer sep=0pt,anchor=north,yshift=-5pt] (B1) at (B.south) {\includegraphics[width=1.2in]{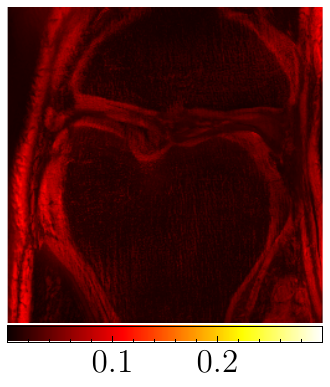}};
			\node[inner sep=0pt, outer sep=0pt,anchor=north,yshift=-5pt] (C1) at (C.south) {\includegraphics[width=1.2in]{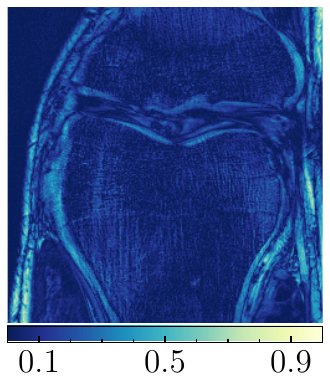}};		
			\node[inner sep=0pt, outer sep=0pt,anchor=north,yshift=-5pt] (X2) at (X1.south) {\includegraphics[width=1.2in]{scan_7.pdf}};
			\node[inner sep=0pt, outer sep=0pt,anchor=north,yshift=-5pt] (A2) at (A1.south) {\includegraphics[width=1.2in]{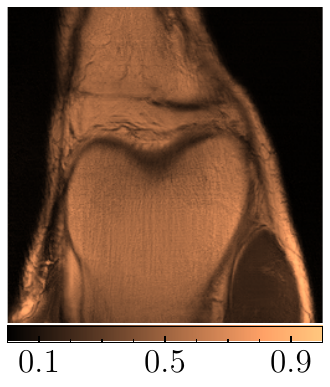}};
			\node[inner sep=0pt, outer sep=0pt,anchor=north,yshift=-5pt] (B2) at (B1.south) {\includegraphics[width=1.2in]{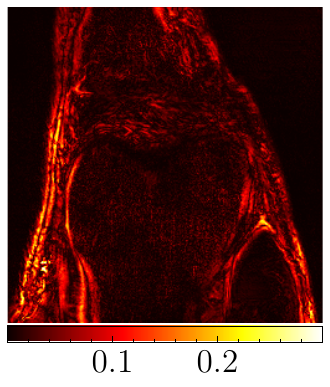}};
			\node[inner sep=0pt, outer sep=0pt,anchor=north,yshift=-5pt] (C2) at (C1.south) {\includegraphics[width=1.2in]{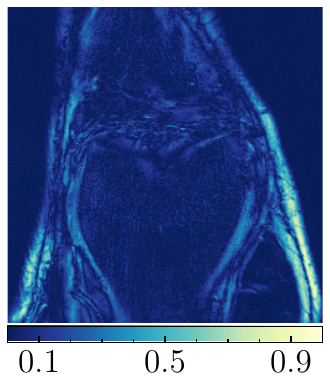}};
			\node[anchor=south] at (X.north) {Ground truth};
			\node[anchor=south] at (A.north) {Post. mean};
			\node[anchor=south] at (B.north) {Post. std. dev.};
			\node[anchor=south] at (C.north) {Abs. error};
			\node[anchor=south,rotate=90,yshift=2pt] at (X.west) {Test slice 1};
			\node[anchor=south,rotate=90,yshift=2pt] at (X1.west) {Test slice 2};
			\node[anchor=south,rotate=90,yshift=2pt] at (X2.west) {Test slice 3};
		\end{tikzpicture}
		\caption{Posterior pixel-wise standard mean (second column) and posterior pixel-wise standard deviation (third column) for various test cases when acceleration factor $r=8$. The first column shows the ground truth slices for reference. The last column shows the absolute error between the posterior mean and the ground truth}
		\label{fig:phase_recon_accel_8}	
	\end{figure}

	The normalizing flow model for this problem consists of 16 affine-coupling flow layers with activation normalization~\cite{kingma2018glow}. See \Cref{appsubsec:phase_nvp} for more details about the scale and shift networks of the affine coupling layers. In this example, we train the normalizing flow model for 50,000 epochs and a batch size of 32. We train the normalizing flow model for every combination of test slice and undersampling mask. Subsequently, for each combination of test slice and mask, we obtain 10,000 samples from the latent posterior distribution to estimate the posterior pixel-wise mean and standard mean. \Cref{fig:phase_recon_accel_4,fig:phase_recon_accel_8} show the posterior mean, posterior standard-deviation, and the absolute error of the posterior mean reconstruction for the four- and eight-fold acceleration, respectively. We compute the  RMSE and SSIM between the posterior mean reconstruction and the ground truth knee slices and report these values in \Cref{tab:phase_recon_metrics}. From \Cref{fig:phase_recon_accel_4,fig:phase_recon_accel_8} and  \Cref{tab:phase_recon_metrics}, we observe that the reconstruction is satisfactory. However, the reconstruction of test slice 1 is comparatively better than those of test slices 2 and 3.  This indicates that the reconstruction of some knee slices, like test slice 1, which probably lies in the typical set of the generator's latent space (range of $G^\ast$~\cite{bora2017compressed}), can be better than atypical test slices. Moreover, for test slice 2, the posterior pixel-wise standard deviation when the acceleration factor $r=4$ is marginally larger in comparison to the case when $r=8$; see the second row third column of \Cref{fig:phase_recon_accel_4,fig:phase_recon_accel_8}. The result above needs further investigation since a reduction in uncertainty as the number of measurements reduces is counter-intuitive. 
	
	Significantly, this example showcases the efficacy of GAN-Flow compared to methods involving MCMC simulations. In this study, we attempted to carry out inference using GAN-HMC. However, we were unsuccessful in obtaining convergence of the Markov chains after varying the number of leapfrog steps and burn-in time. We posit that this may be due to the relatively large latent dimensionality in this study.
	\begin{table}[H]
		\centering
		\caption{RMSE and SSIM of the of the posterior mean reconstruction of the test slices for different accelerations}
		\label{tab:phase_recon_metrics}
		\begin{tabular}{@{\extracolsep{2pt}}l cc cc}
			\toprule[1.5pt]
			\multirow{2}{*}{Test slice}& \multicolumn{2}{c}{4$\times$ acceleration} & \multicolumn{2}{c}{8$\times$ acceleration}\\
			\cline{2-3}\cline{4-5}
			& RMSE& SSIM& RMSE& SSIM\\
			\midrule[1.5pt]
			Slice 1& 0.0070& 0.6185& 0.0072& 0.6142\\
			Slice 2& 0.0238& 0.4802& 0.0246& 0.4587\\
			Slice 3& 0.0206 & 0.5556& 0.0217& 0.5702\\
			\bottomrule[1.5pt]
		\end{tabular}
	\end{table}

	\section{Conclusions}\label{sec:conclusion}
	
	Bayesian inference is widely applicable but its application is challenging, especially, in cases where the inverse problem is high-dimensional and prior information is qualitative in nature. In this work, we propose GAN-Flow, a dimension-reduced variational approach to solving large-scale inverse problems. GAN-Flow combines together two types of generative models --- GANs and normalizing flows. The former is used to form an informative data-driven prior with the generator providing a map between a low-dimensional latent space and the high-dimensional ambient space. Normalizing flows are used to solve the inverse problem variationally in the low-dimensional latent space, made possible due to an invertible map that transforms the prior distribution in the latent space into the posterior distribution in the latent space. The low-dimensional latent posterior can be sampled, without evaluating the underlying forward model, to obtain samples from the high-dimensional ambient space. We have also shown how GAN-Flow can be used to estimate statistics of the ambient posterior distribution. We have used GAN-Flow to solve three physics-based inverse problems that include various challenging scenarios. In particular, the application to phase imaging shows that GAN-Flow can handle challenging prior information, nonlinear forward models, and very large-scale inverse problems. The extension of GAN-Flow to black-box forward models that are incompatible with automatic differentiation is a promising research direction that will make GAN-Flow more widely applicable. 

	\section{Acknowledgments}
	
	This work was carried out while the first author was a graduate student in the Sonny Astani Department of Civil \& Environmental Engineering at the University of Southern California.  Agnimitra Dasgupta acknowledges support from the University of Southern California through a Provost's Ph.D. Fellowship.  Agnimitra Dasgupta and Erik Johnson also gratefully acknowledge the support of this work by the National Science Foundation through award CMMI 16-63667. Assad A Oberai gratefully acknowledges support from ARO, USA grant W911NF2010050. Any opinions, findings, and conclusions or recommendations expressed in this material are those of the authors and do not necessarily reflect the views of NSF or USC. The authors also acknowledge the Center for Advanced Research Computing (CARC, \href{https://carc.usc.edu}{carc.usc.edu}) at the University of Southern California for providing computing resources that have contributed to the research results reported within this publication.
	
	\bibliographystyle{elsarticle-num-names} 
	\bibliography{references}
	
	\newpage
	\appendix
	\setcounter{section}{0}
	\renewcommand{\appendixname}{Appendix}
	\renewcommand{\thesection}{Appendix~\Alph{section}}
	\renewcommand{\thesubsection}{\Alph{section}\arabic{subsection}}
	\renewcommand{\thefigure}{\Alph{section}\arabic{figure}}
	\setcounter{figure}{0}
	\renewcommand{\thetable}{\Alph{section}\arabic{table}}
	\setcounter{table}{0}

	\section{Details of the GAN and normalizing flow architectures of GAN-Flow and training hyper-parameters}\label{appsubsec:models}
	In this section we describe the WGAN-GP and normalizing flow models used in the GAN-Flow pipeline for various inverse problems. Some of the nomenclature we use are as follows:
	\begin{enumerate}[itemsep=-5pt]
		\item FC($n$) --- Fully connected layer of width $n$.
		\item Tr. Conv2D ($c_{\mathrm{out}}$, $k$, $s$, $p$, $p_{\mathrm{out}}$) --- 2D transpose convolution layer with $c_{\mathrm{out}}$ output channels, kernel size $(k,k)$, stride $s$, padding $p$ and output padding $p_{\mathrm{out}}$.
		\item Conv2D ($c_{\mathrm{out}}$, $k$, $s$, $p$) --- 2D convolution layer $c_{\mathrm{out}}$ output channels, kernel size $(k,k)$, stride $s$ and padding $p$.
		\item Self Attention --- self-attention module~\cite{zhang2019self}.
		\item BN, LN, PixelNorm, Mini-batch std. dev. normalization  --- batch, layer, pixel~\cite{karras2018progressive} and mini-batch standard deviation~\cite{karras2018progressive} normalization, respectively.
		\item LReLU($\alpha$), ELU, and TanH --- Leaky rectified linear unit (with parameter $\alpha$), exponential linear unit and hyperbolic tangent activation functions, respectively.
		\item Up-sample 2$\times$ -- Up-scaling by a factor of 2 using bi-linear interpolation.
		\item  Down-sample 2$\times$ -- Average pooling over 2$\times$2 patches with stride 2. 
	\end{enumerate}
		
	\subsection{WGAN-GP model architectures}\label{appsubsec:gan_models}
\begin{figure}[t]
		\centering
		\begin{tikzpicture}
			\node[draw=red] (A) at (0,0) {\includegraphics[height=2in]{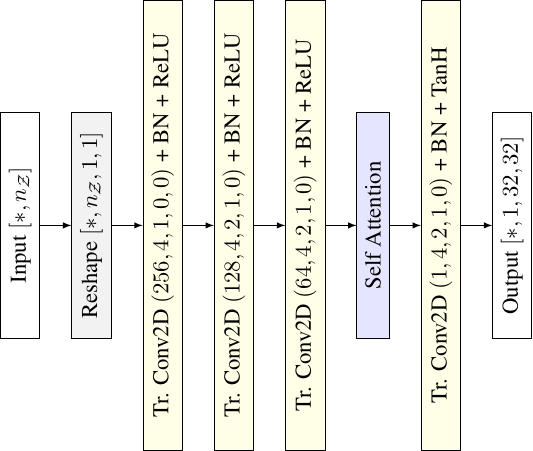}};
			\node[draw=blue, anchor=west,xshift=40pt] (B) at (A.east) {\includegraphics[height=2in]{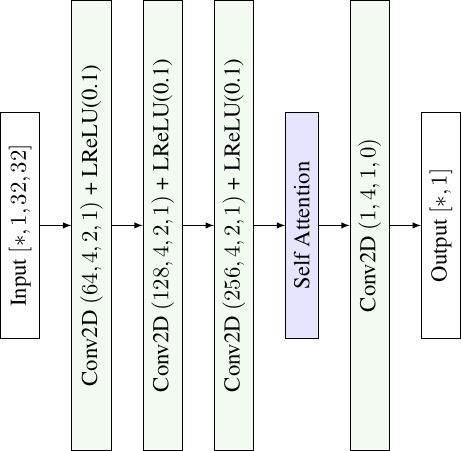}};
			\node[inner sep=0pt,anchor=north,xshift=-10pt, yshift=-5pt] at (A.north west) {(a)};
			\node[inner sep=0pt,anchor=north,xshift=-10pt, yshift=-5pt] at (B.north west) {(b)};
			\node[anchor=north,align=center,color=red] at (A.south) {Generator};
			\node[anchor=north,align=center,color=blue] at (B.south) {Critic};
		\end{tikzpicture}
		\caption{(a) Generator and (b) critic architectures of the WGAN-GP model for the initial condition inference inverse problem}
		\label{fig:initcond_gan}
	\end{figure}

\begin{figure}[t]
		\centering
		\begin{tikzpicture}
			\node[draw=red] (A) at (0,0) {\includegraphics[height=2in]{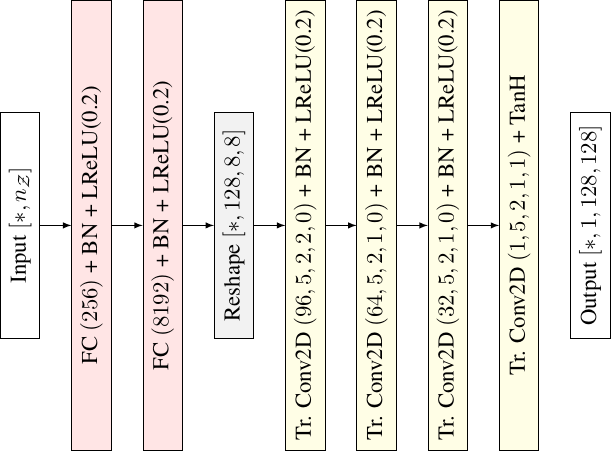}};
			\node[draw=blue, anchor=west,xshift=30pt] (B) at (A.east) {\includegraphics[height=2in]{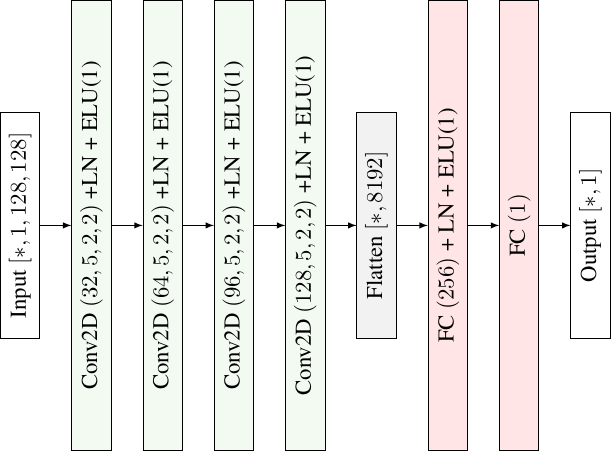}};
			\node[inner sep=0pt,anchor=north,xshift=-10pt, yshift=-5pt] at (A.north west) {(a)};
			\node[inner sep=0pt,anchor=north,xshift=-10pt, yshift=-5pt] at (B.north west) {(b)};
			\node[anchor=north,align=center,color=red] at (A.south) {Generator};
			\node[anchor=north,align=center,color=blue] at (B.south) {Critic};
		\end{tikzpicture}
		\caption{(a) Generator and (b) critic architectures of the WGAN-GP model for the inverse Radon transform problem}
		\label{fig:tomo_gan}
	\end{figure}

	\Cref{tab:model_details} lists the training hyper-parameters. \Cref{fig:initcond_gan} shows the generator and critic architecture used in the initial condition inference problem. \Cref{fig:tomo_gan} shows the generator and architecture used in the inverse Radon transform problem. \Cref{fig:phase_gan}(a)~and~(c) shows the generator and critic architectures, respectively, we use for the phase retrieval problem. The generator comprises convolution blocks that are shown in \Cref{fig:phase_gan}(b). Similarly, the critic is made of a convolution block that is denoted as `Dis. Convolution Block' in \Cref{fig:phase_gan}(c) and shown in \Cref{fig:phase_gan}(b). In this study, we use the Progressive growing of GAN methodology to train the generator and critic networks. We briefly summarize the ProGAN method here, and refer \cite{karras2018progressive} to interested readers for more details. 
	\begin{table}[H]
		\small
		\centering
		\caption{Training hyper-parameters for WGAN-GP and normalizing flow models of the GAN-Flow pipeline}
		\label{tab:model_details}
		\begin{tabular}{l l ccc}
			\toprule[1.5pt]
			\multirow{2}{*}{Model} & \multirow{2}{*}{hyper-parameter}& \multicolumn{3}{c}{Inverse problem} \\
			\cline{3-5}
			& & \makecell{Heat conduction\\ (\Cref{subsec:heat})} & \makecell{Radon transform\\(\Cref{subsec:tomo})} & \makecell{Phase imaging\\ (\Cref{subsec:phase})}  \\
			\midrule[1.5pt]
			\parbox[b]{2mm}{\multirow{8}{*}{\rotatebox[origin=c]{90}{Wasserstein GAN}}}&Latent dimension $\Nz$ &5&60&512\\
			&Architecture&\Cref{fig:initcond_gan}&\Cref{fig:tomo_gan}&\Cref{fig:phase_gan}\\
			&Training epochs&500&1000&294\\
			&Learning rate&0.0002&0.001&0.003\\
			&Gradient penalty $\lambda$&10&10&10\\
			&Batch size&64&100& 128 $\to$ 64\\
			&$n_{\mathrm{critic}}/n_{\mathrm{gen}}$&5&4&1\\
			&Optimizer& \makecell{Adam \\ $\beta_1 = 0, \beta_2 = 0.99$}&\makecell{Adam \\ $\beta_1 = 0.5, \beta_2 = 0.99$}&\makecell{Adam \\ $\beta_1 = 0, \beta_2 = 0.99$}\\
			\midrule
			\parbox[b]{2mm}{\multirow{6}{*}{\rotatebox[origin=c]{90}{Normalizing flow}}}&Type of flow model layer& Planar& Planar&Affine coupling\\
			&Number of flow layers $n_{\mathrm{f}}$& 64&256&16\\
			&Training epochs& 5000&15000&50000\\
			&Learning rate& 0.002&0.002&0.001\\
			&Batch size& 32&32&32\\
			&Optimizer& \makecell{Adam \\ $\beta_1 = 0.9, \beta_2 = 0.999$}&\makecell{Adam \\ $\beta_1 = 0.9, \beta_2 = 0.999$}&\makecell{Adam \\ $\beta_1 = 0.9, \beta_2 = 0.999$}\\
			\bottomrule[1.5pt]
		\end{tabular}
	\end{table}	
	

	\subsubsection{Progressive growing of GANs}
\begin{figure}[t]
		\centering
		\begin{tikzpicture}
			\node[draw=red, anchor=west] (A) at (0,0) {\includegraphics[height=2.33in]{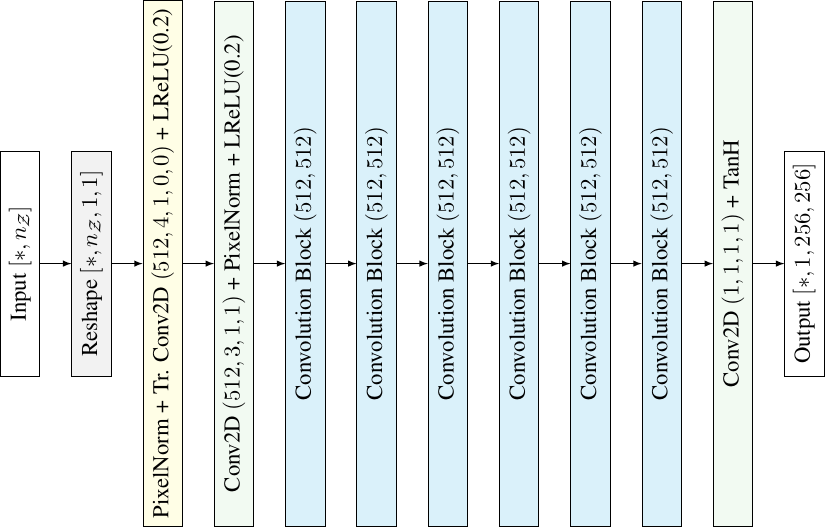}};
			
			\node[draw=gray, anchor=west] (B) at (12,0) {\includegraphics[height=2.33in]{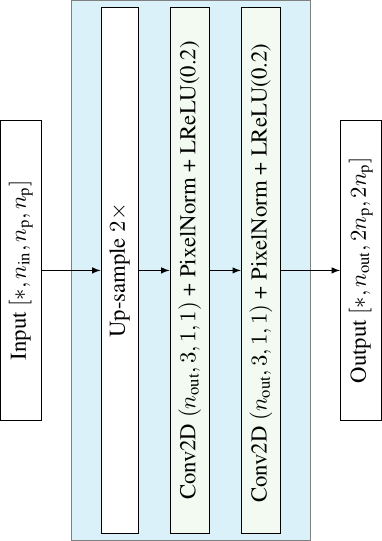}};
			
			\node[draw=blue,anchor=west] (C) at (0, -7) {\includegraphics[height=2.33in]{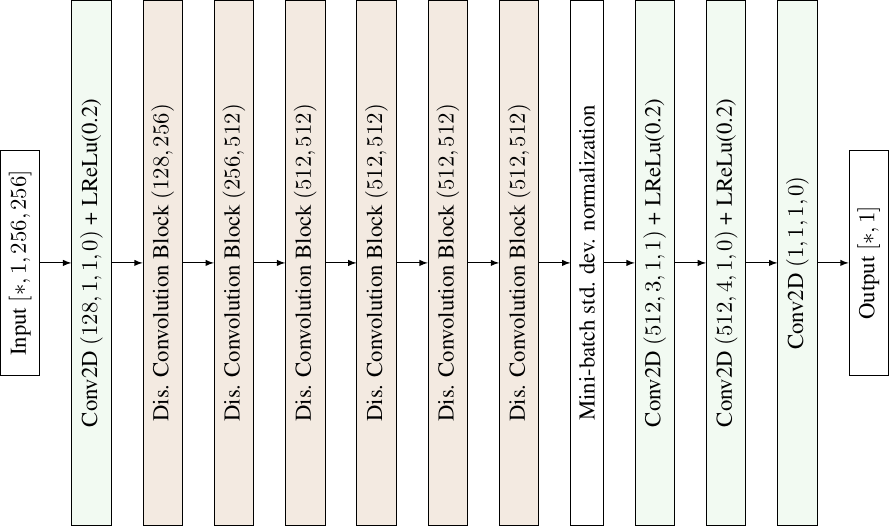}};
			
			\node[draw=gray,anchor=west] (D) at (12, -7) {\includegraphics[height=2.33in]{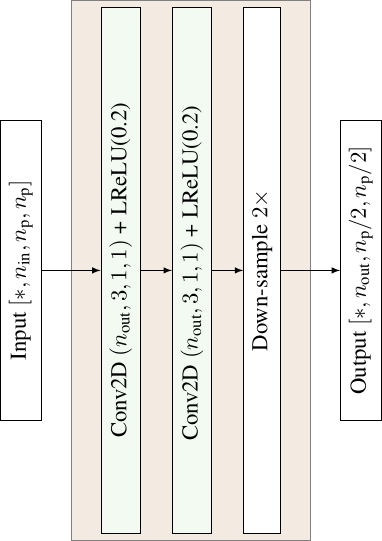}};
			
			\node[inner sep=0pt,anchor=north,xshift=-10pt, yshift=-5pt] at (A.north west) {(a)};
			\node[inner sep=0pt,anchor=north,xshift=-10pt, yshift=-5pt] at (B.north west) {(b)};
			\node[inner sep=0pt,anchor=north,xshift=-10pt, yshift=-5pt] at (C.north west) {(c)};
			\node[inner sep=0pt,anchor=north,xshift=-10pt, yshift=-5pt] at (D.north west) {(d)};
			
			\node[anchor=north,align=center,color=red] at (A.south) {Generator};
			\node[anchor=north,align=center,color=blue] at (C.south) {Critic};
			\node[anchor=north,align=center,color=black] at (B.south) {Convolution Block};
			\node[anchor=north,align=center,color=black] at (D.south) {Dis. Convolution Block};
			
		\end{tikzpicture}
		\caption{(a) Generator and (c) critic architectures of the WGAN-GP model for the phase retrieval problem. (b) and (d) shows the convolution blocks within the generator and critic, respectively}
		\label{fig:phase_gan}
	\end{figure}

\begin{figure}[t]
		\begin{tikzpicture}
			\node[inner sep=0pt, outer sep=0pt,] (A) at (0,0) {\includegraphics[height=1.5in]{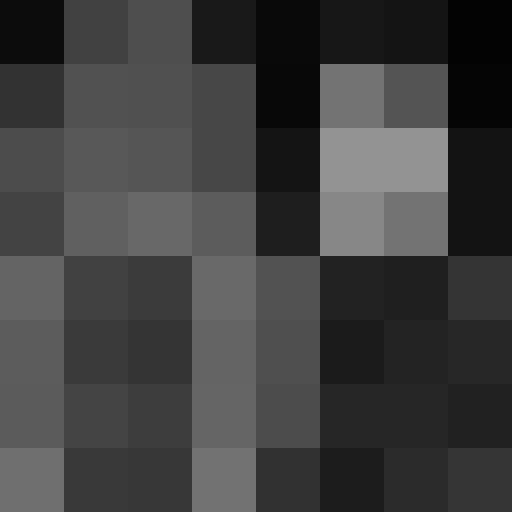}};
			\node[inner sep=0pt, outer sep=0pt,anchor=west, xshift=10pt] (B) at (A.east) {\includegraphics[height=1.5in]{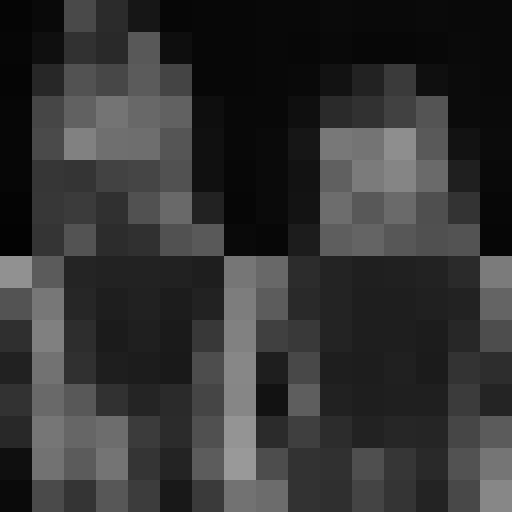}};
			\node[inner sep=0pt, outer sep=0pt,anchor=west, xshift=10pt] (C) at (B.east) {\includegraphics[height=1.5in]{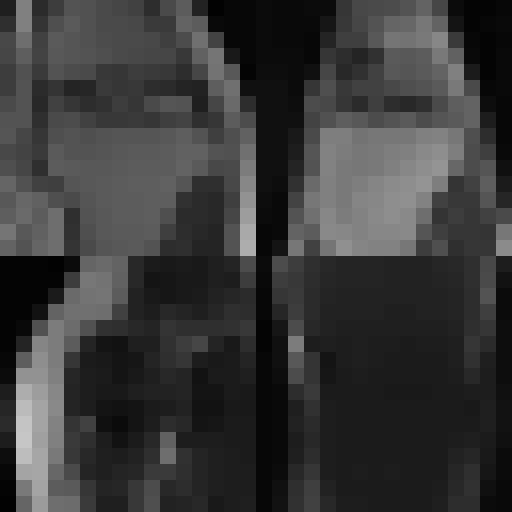}};
			\node[inner sep=0pt, outer sep=0pt,anchor=west, xshift=10pt] (D) at (C.east) {\includegraphics[height=1.5in]{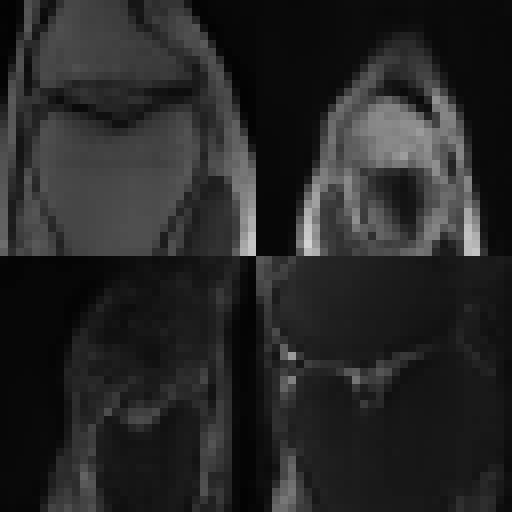}};
			
			\node[inner sep=0pt, outer sep=0pt,anchor=north, yshift=-50pt] (E) at (A.south) {\includegraphics[height=1.5in]{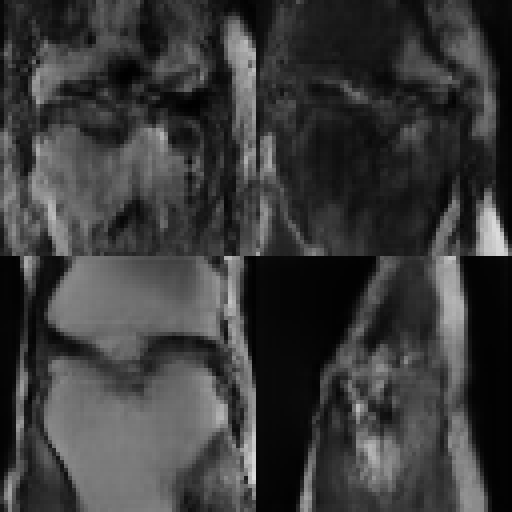}};
			\node[inner sep=0pt, outer sep=0pt,anchor=west, xshift=10pt] (F) at (E.east) {\includegraphics[height=1.5in]{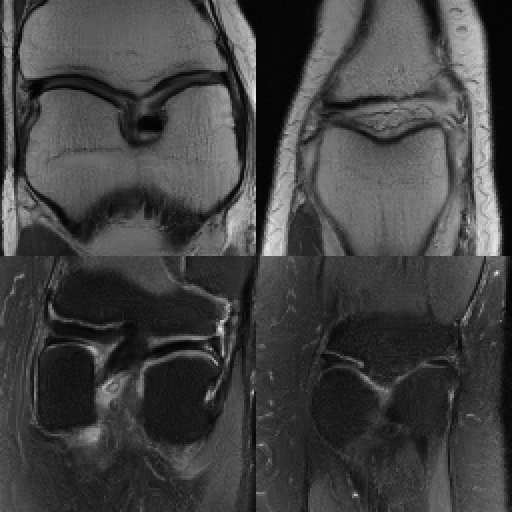}};
			\node[inner sep=0pt, outer sep=0pt,anchor=west, xshift=10pt] (G) at (F.east) {\includegraphics[height=1.5in]{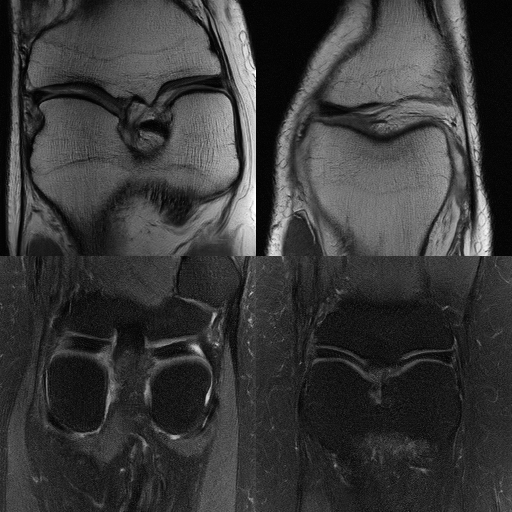}};
			
			\node[align=center,anchor=south] at (A.north) {Stage 2\\Image size 4$\times$4};
			\node[align=center,anchor=south] at (B.north) {Stage 3\\Image size 8$\times$8};
			\node[align=center,anchor=south] at (C.north) {Stage 4\\Image size 16$\times$16};
			\node[align=center,anchor=south] at (D.north) {Stage 5\\Image size 32$\times$32};
			\node[align=center,anchor=south] at (E.north) {Stage 6\\Image size 64$\times$64};
			\node[align=center,anchor=south] at (F.north) {Stage 7\\Image size 128$\times$128};
			\node[align=center,anchor=south] at (G.north) {Stage 8\\Image size 256$\times$256};		
		\end{tikzpicture}
		\caption{Knee slices of resolution 2\textsuperscript{$k$}$\times$2\textsuperscript{k} generated at the end of the \supth{k} training stage using the WGAN-GP model for the phase retrieval problem}
		\label{fig:phase_prior_stages}
	\end{figure}

	In the ProGAN methodology, both generator and critic are trained synchronously to synthesize images starting from size  4$\times$4 up until 256$\times$256. For instance, at the first stage, when the GAN is learning to synthesize images of size 4$\times$4, only the first three layers (after the inputs are reshaped and including the first convolution block) of the generator is trained. Similarly, at this stage, the discriminator only consists of first two layers (including the first Dis.~Convolution block) and the final four layers (starting from the mini-batch standard deviation normalization). After the first stage of training is complete, a new convolution block is now appended to the previously trained convolution block in the generator. Similarly, another `Dis.~Convolution Block' is appended to the previously trained block of similar type to the discriminator. All weights of the generator and critic networks are updated at this stage. This process continues for a further four stages, thus a total six stages, up until the GAN learns to synthesize images of the required size. The requisite number of stages $n_\mathrm{stage}$ should satisfy $2^{n_\mathrm{stage}+1}=128$. In this study, the desired size of the knee slices was 256$\times$256, which necessitates $n_\mathrm{stage}=7$. We train the GAN for 42 epochs at each stage, which makes a total of 294 epochs across all stages. We use a batch size of 128 for the first four stages, and reduce it by half for the penultimate three stages. We also randomly flip the images horizontally to augment the training dataset. At every stage lower-resolution images from the prior dataset are down-sampled using average pooling to obtain the necessary `\emph{real}' images for training. Moreover, during training at the \supth{k} stage ($k$ starts from 2 going to 8 corresponding to resolutions of 4$\times$4 to 128$\times$128) beyond $k=2$, the synthesized images are formed by a linear combination of the up-sampled images from the previous generator up to the previous stage and the current stage using residual connections. Similarly, the critic also blends together image features at the \supth{(k-1)} resolution level using residual connections. This linear superposition factor, say $\alpha$, linearly increases between 0 and 1 through the training epochs to ultimately only consider the images entirely synthesized at the \supth{k} resolution level, \ie the contribution from the residual connections gradually fades away as training progresses for every stage. Figure 2 in \cite{karras2018progressive} is instructive of this multi-scale blending. \Cref{fig:phase_prior_stages} shows knee slices of various resolutions generated at the end of \supth{k} stage of training using the WGAN-GP model from \Cref{fig:phase_gan}. 
	
	\subsection{Normalizing flow model architectures}\label{appsubsec:phase_nvp}
\begin{figure}[t]
		\centering
		\begin{tikzpicture}
			\node[draw=green!40!black!80] (A) at (0,0) {\includegraphics[height=2in]{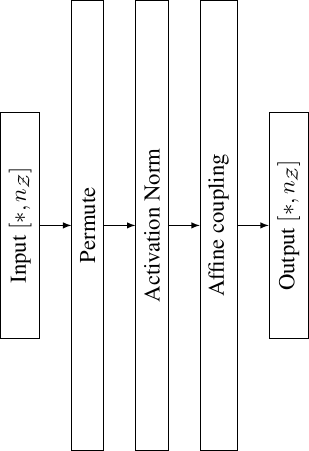}};
			\node[draw=green!40!black!80, anchor=west,xshift=30pt] (B) at (A.east) {\includegraphics[height=2in]{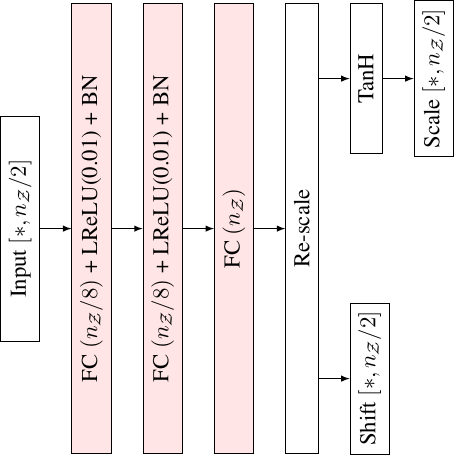}};
			\node[inner sep=0pt,anchor=north,xshift=-10pt, yshift=-5pt] at (A.north west) {(a)};
			\node[inner sep=0pt,anchor=north,xshift=-10pt, yshift=-5pt] at (B.north west) {(b)};
		\end{tikzpicture}
		\caption{(a) Typical flow layer with affine coupling transform, and (b) scale and shift networks used for the affine coupling transform in the phase retrieval problem}
		\label{fig:phase_nvp}
	\end{figure}

	We use planar flow layers with hyperbolic tangent non-linearity for both the initial condition inference and inverse Radon transform problem. For the phase retrieval problem, we use affine coupling flow layers, as shown in \Cref{fig:phase_nvp}(a), and the scale and shift networks are shown in \Cref{fig:phase_nvp}(b). Every affine coupling block permutes its input such that the partition, described above \Cref{eq:nvp_split}, is random for every layer; this promotes better mixing among every latent dimension. Subsequently, activation normalization is applied, which scales the inputs to have zero mean and unit variance; this transformation is also updated during training. The re-scaling layer in \Cref{fig:phase_nvp}(b) has a single learnable parameter that simply scales and multiplies itself with the output from the previous layer. This parameter is initially set to zero such that the whole layer starts out as an identity transform. After the re-scaling layer, TanH operation operates on one-half of the data and serves as the scale operator, while the other half acts as the shift operator.
	
	\section{Additional results}\label{appsubsec:addexp}
	\subsection{initial condition inference}\label{appsubsec:initcond_addexp}
\begin{figure}
		\centering
		\begin{tikzpicture}
			\node[inner sep=0pt,outer sep=0pt] (A) at (0,0) {\includegraphics[height=1.7in]{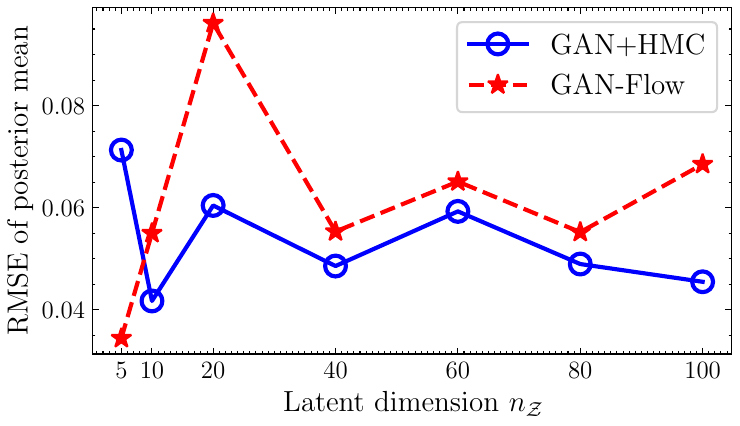}};
			\node[anchor=west, xshift=25pt, inner sep=0pt,outer sep=0pt] (B) at (A.east) {\includegraphics[height=1.7in]{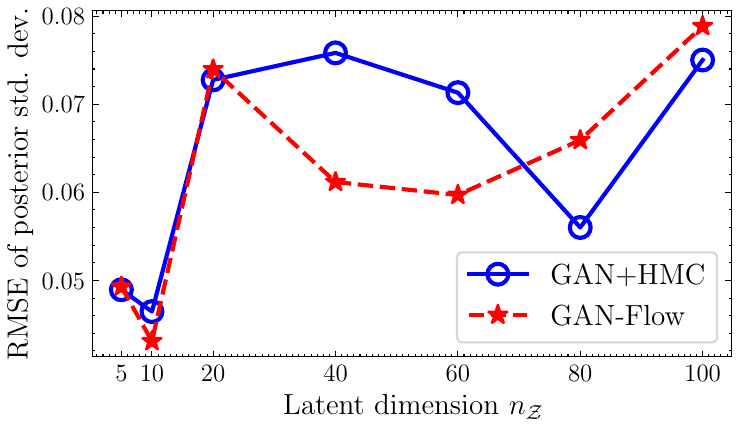}};
			\node[inner sep=0pt,anchor=north,xshift=-10pt, yshift=-5pt] at (A.north west) {(a)};
			\node[inner sep=0pt,anchor=north,xshift=-10pt, yshift=-5pt] at (B.north west) {(b)};
		\end{tikzpicture}
		\caption{RMSE of the posterior pixel-wise (a) mean and (b) standard-deviation estimated using various methods with respect to the corresponding statistics estimated using MCS for the initial condition inference problem}
		\label{fig:initcond_varyNz}
	\end{figure}

	For the initial condition inference inverse problem, we vary the latent dimensionality $\Nz$, while keeping all other hyper-parameters fixed, to study its effect on the overall performance of GAN-Flow. \Cref{fig:initcond_varyNz} reveals that the performance of both GAN-Flow and GAN+HMC deteriorates as the latent dimensionality $\Nz$ increases. One reason for this may be the following: the latent space dimensionality controls the expressivity of the generator, and a larger than necessary latent space dimensionality may be introducing spurious uncertainty in the prior. Additionally, the deteriorating performance can also be attributed to the curse of dimensionality. As the latent space dimensionality increases, both HMC and normalizing flows find it increasingly harder to sample the latent posterior distribution. In a nutshell, \Cref{fig:initcond_varyNz} is empirical evidence of the fact that dimension reduction is beneficial for Bayesian inference. From \Cref{fig:initcond_varyNz}, $\Nz=5$ leads to the lowest RMSE in the estimated posterior statistics. This is expected since the underlying prior distribution has only four random variables.  While it may appear from \Cref{fig:initcond_varyNz}(a) that GAN-Flow is not able to estimate the posterior mean as well as GAN+HMC, recall that we train the normalizing flows with the same hyper-parameter setting and a total of 3.2$\times$10\textsuperscript{4} forward model evaluations irrespective of the value of $\Nz$. Increasing the computational effort as $\Nz$ increases should help improve the performance of GAN-Flow. In \Cref{fig:initcond_varyNz}(b), GAN-Flow does at least as well as GAN+HMC in capturing the posterior variance at modest dimensions ($\Nz < 80$). 
	
	\subsection{Inverse Radon transform}\label{appsubsec:tomo_addexp}
	
\begin{figure}[H]
		\centering
		\begin{tikzpicture}
			\node[inner sep=0pt,outer sep=0pt] (A) at (0,0) {\includegraphics[height=1.8in]{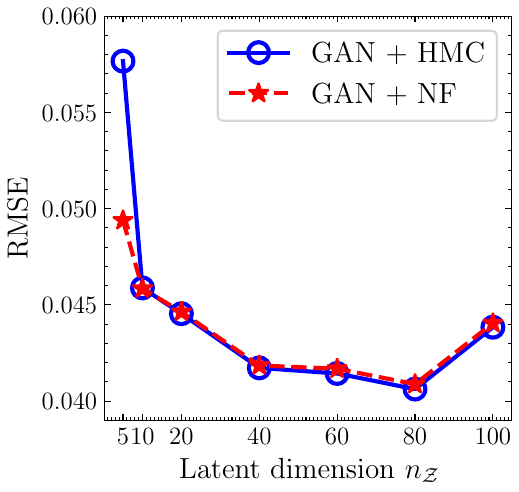}};
			\node[anchor=west, xshift=25pt, inner sep=0pt,outer sep=0pt] (B) at (A.east) {\includegraphics[height=1.8in]{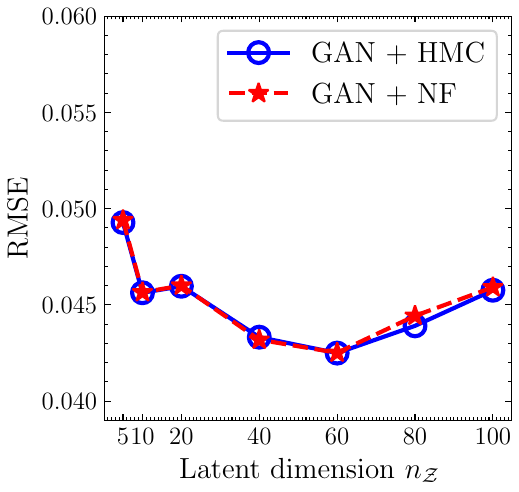}};
			\node[anchor=west, xshift=25pt, inner sep=0pt,outer sep=0pt] (C) at (B.east) {\includegraphics[height=1.8in]{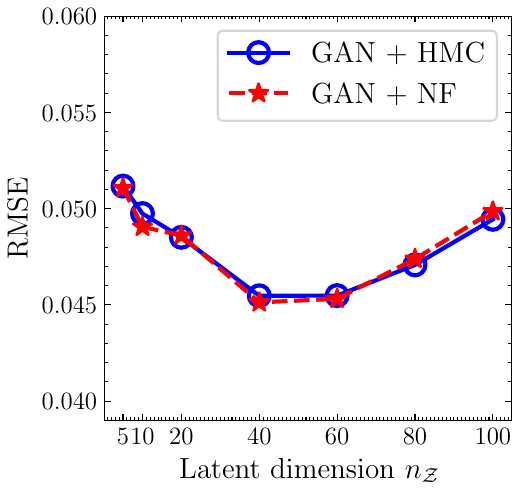}};
			
			\node[inner sep=0pt,anchor=south,outer sep=0pt,xshift=10pt] at (A.north) {$\sigma_{\eta}^2 = 1$};
			\node[inner sep=0pt,anchor=south,outer sep=0pt,xshift=10pt] at (B.north) {$\sigma_{\eta}^2  = 10$};
			\node[inner sep=0pt,anchor=south,outer sep=0pt,xshift=10pt] at (C.north) {$\sigma_{\eta}^2  = 50$};
			
		\end{tikzpicture}
		\caption{RMSE of the posterior pixel-wise mean with respect to the `\emph{true}' phantom for varying levels of variance $\sigma^2_{\eta}$ of the measurement noise in the inverse Radon transform problem}
		\label{fig:tomo_varyNz}
	\end{figure}

	Similar to the previous example, we vary the latent dimensionality $\Nz$, while keeping all other hyper-parameters fixed, to study its effect on the overall performance of GAN-Flow across different levels of measurement noise. We plot the reconstruction error of the posterior mean in \Cref{fig:tomo_varyNz}, which shows that the optimal latent dimension $\Nz$ lies between 40 and 80 for the three levels of measurement noise we consider. Note that, in this example, the underlying prior distribution is parameterized by 13 variables. Hence, the reconstructions from GAN-Flow and GAN-HMC are inadequate when $\Nz \leq 20$. At such low latent dimensions, the prior distribution from the WGAN-GP is not sufficiently expressive. Beyond that, the reconstruction error first reduces and then increases again. The comparatively larger reconstruction errors when $\Nz > 80$ may be due to insufficient training of the normalizing flow or the inefficacy of HMC in sampling high-dimensional distributions. Therefore, for this example, we choose $\Nz = 60$ to obtain a balanced performance from GAN-Flow across all levels of measurement noise.  
	
\end{document}